\documentclass[submission,copyright,creativecommons]{eptcs}
\providecommand{\event}{Master of Science in Logic, Computation, and Methodology} 

\usepackage{iftex}
\usepackage{graphicx}
\usepackage{lipsum}
\usepackage{amsmath}
\usepackage{amssymb}
\usepackage{proof} 
\ifpdf
  \usepackage{underscore}         
  \usepackage[T1]{fontenc}        
\else
  \usepackage{breakurl}           
\fi

\title{Topological Semantics for Common \\ Inductive Knowledge}
\author{Siddharth Namachivayam
\institute{Department of Philosophy \\
Carnegie Mellon University\thanks{The author would like to thank Adam Bjorndahl and Kevin Kelly for advising this project. Preliminary progress was presented in 2025 at the Indian Statistical Institute's conference on Recent Trends in Logic and Game Theory as well as the Progic Conference Series' 12th Workshop on Combining Logic and Probability. This work also received financial support in the summer of 2024 from Carnegie Mellon University's Institute for Complex Social Dynamics.}\\
Pittsburgh, Pennsylvania}
\email{snamachi@andrew.cmu.edu}}
\def\titlerunning{Topological Semantics for Common Inductive Knowledge}
\def\authorrunning{Siddharth Namachivayam}
\begin{document}
\begin{titlepage}
\centering

\vspace*{1.5in}

{\LARGE \textbf{Topological Semantics for Common Inductive Knowledge}\par}

\vspace{0.2in}

{\large Siddharth Namachivayam\par}

\vspace{0.2in}

{\large April 2026\par}

\vspace{1in}

{\large Department of Philosophy\par}
\vspace{0.2in}
{\large Carnegie Mellon University\par}
\vspace{0.2in}
{\large Pittsburgh, PA 15213\par}

\vspace{0.5in}

{\large \textbf{Thesis Committee:}\par}
\vspace{0.2in}
{\large Prof.\ Adam Bjorndahl\par}
\vspace{0.2in}
{\large Prof.\ Kevin Kelly\par}

\vspace{1in}

{\itshape
Submitted in fulfillment of the requirements for the degree:\par
\vspace{0.2in}
Master of Science in Logic, Computation, and Methodology
\par}

\vfill

\end{titlepage}
\tableofcontents
\maketitle
\begin{abstract}
Consider a community of scientists whose labs are each capable of conducting a different set of experiments. The scientists want to work together to confirm a new hypothesis, but to ensure blindness, their labs generally prohibit the scientists from communicating with each other. Further, each scientist can only make so many retractions to their lab before having to cease inquiry and suspend judgement forever. How might the scientists coordinate whether to affirm or suspend judgement on this hypothesis in light of their private experiments so that their labs are guaranteed to converge to the same conclusion and that this conclusion will not be a false positive? Call this problem `inductive coordinated attack.' In this paper, we develop a logic for solving inductive coordinated attack by determining when and how a hypothesis can become what we call `common inductive knowledge.' We begin by precisifying Lewis’ account of common knowledge in \emph{Convention} which describes the generation of higher-order expectations between agents as hinging upon agents’ inductive standards and a shared witness.  Our language has a rather rich syntax in order to capture equally rich notions central to Lewis’ account; for instance, we speak of an agent ‘having inductive reason to believe’ a proposition and one proposition ‘indicating’ to an agent that another proposition holds. This syntax affords a novel topological semantics which, following Kelly 1996's approach in \emph{The Logic of Reliable Inquiry}, takes as primitives agents' information bases. In particular, we endow each agent with a `switching tolerance' meant to represent their personal inductive standards for learning. After establishing soundness of our proof system with respect to this semantics, we conclude by showing how our logic can be used to solve inductive coordinated attack.
\end{abstract}

\section{Introduction}

\noindent In \emph{Convention}, David Lewis presents 3 conditions for a witness $W$ to generate common knowledge of a proposition $P$.\footnote[1]{Strictly speaking, Lewis defines what it means for $W$ to be a `basis' for common knowledge of $P$. We will instead call $W$ a `witness,' reserving the term `basis' to be used in its topological sense.} These are:
\begin{enumerate}
\item Everyone has reason to believe $W$.
\item $W$ indicates to everyone that $P$.
\item $W$ indicates to everyone that 1.
\end{enumerate}
where $W$ indicates to an agent $i$ that $P$ iff:
\begin{quote}
\centering
if $i$ had reason to believe $W$, $i$ would thereby have reason to believe $P$.	
\end{quote}
For Lewis, what $W$ indicates to $i$ (and whether $W$ generates common knowledge of $P$) crucially depends on $i$'s inductive standards.  After making appropriate `ancillary' assumptions about the nature of these inductive standards, Lewis uses his 3 conditions to derive that:
\begin{enumerate}
\item[$(1)$] Everyone has reason to believe $P$.
\item[$(2)$] Everyone has reason to believe $(1)$.
\item[$(3)$] Everyone has reason to believe $(2)$.
\item[] ... \ \ \ \ \ ... \ \ \ \ \ ... \ \ \ \ \ ... \ \ \ \ \ ... \ \ \ \ \ ... \ \ \ \ \ ... \ \ \ \ \ ...
\item[$(n)$] Everyone has reason to believe $(n-1)$.
\end{enumerate}
However, Lewis is rather opaque about what his `ancillary' assumptions exactly are. Understandably, subsequent formal treatments of common knowledge largely tended to overlook this critical inductive dimension in their accounts. Additionally, most formal treatments have also neglected the role of witnesses in generating common knowledge. For instance, at a pre-theoretic level, it is widespread in the game theory literature to speak of common knowledge/belief of $P$ and mean no more than the obtaining of some infinite collection of propositions like:
\begin{enumerate}
\item[$(1)'$] Everyone knows/believes $P$.
\item[$(2)'$] Everyone knows/believes $(1)'$.
\item[$(3)'$] Everyone knows/believes $(2)'$.
\item[] ... \ \ \ \ \ ... \ \ \ \ \ ... \ \ \ \ \ ... \ \ \ \ \ ... \ \ \ \ \ ... \ \ \ \ \ ... \ \ \ \ \ ...
\item[$(n)'$] Everyone knows/believes $(n-1)'$.
\end{enumerate}
As a result, the original Lewisian picture of a witness generating common knowledge has largely been swept under the rug. Cubitt \& Sugden 2003 are a notable exception; they attempt to painstakingly reconstruct Lewis' vision and provide an account of what his `ancillary' assumptions about agents' inductive standards might be. While such an exercise is not without its merits, the primary purpose of this paper is not to argue about what Lewis did or did not think. Instead, we will aim to provide a logic for capturing how common inductive knowledge can be generated by a witness that dispenses with the need for Lewis' `ancillary' assumptions as much as possible while still maintaining the spirit of his original picture. 

\section{Pre-Theory}
Sharpening Lewis' view, our primary thesis will be to clearly distinguish the notion of `having \emph{inductive} reason to believe $W$' from the notion of `having $W$ as \emph{deductive} reason to believe $P$.' At a high level, imagine agents receive possibly distinct streams of evidence and have different `switching tolerances' for how many times they are willing to change their mind while limit deciding a proposition. In this setting, when we say an agent $i$ \emph{has} reason to believe a proposition, we intend to convey that there is some (possibly transitory) state of affairs where $i$ is justified in believing this proposition. Namely, we say that agent $i$ has \emph{inductive} reason to believe $W$ iff $i$ can `feasibly' and `efficiently' limit decide $W$ after affirming it in light of some true evidence $E$, where `feasibly' means doing so within agent $i$'s switching tolerance while `efficiently' means doing so with as few mind changes as methodologically possible. This way of modeling inductive justification can be seen as an alternative to Bayesianism, where instead of conditioning a credence function, an agent's primary guiding epistemic norm is to follow Ockham's razor (Kelly 2004). On this view, methods which make unnecessary mind changes are needlessly complex. Hence, an agent should chose to believe $W$ in light of true evidence $E$ only if doing so is part of a `simple' method which minimizes the number times they will have to change their mind. 
\newpage
\noindent Next, we say that $i$ has $W$ as \emph{deductive} reason to believe $P$ iff:
\begin{itemize}
\item $i$ has inductive reason to believe $W$.
\item If $E$ is some true evidence which gives $i$ inductive reason to believe $W$ then $W \land E$ entails $P$. 
\end{itemize}
We take the latter requirement above to constitute the meaning of the phrase `$W$ indicates to $i$ that $P$' since it can be glossed as saying `if $i$ had inductive reason to believe $W$, $i$ would thereby have $W$ as deductive reason to believe $P$.'\footnote[2]{We stress that indication is \emph{not} mere entailment. Nevertheless, it may be helpful for readers to think of indication as just being mere entailment on a first pass of this section.} According to these definitions then, we have as primitive rule of inference:
$$\infer
{\text{ \ \ \ \ \ \ \ \ \ \ \ \ \ \ \ \ \ } i \text{ has $A$ as deductive reason to believe } B \text{ \ \ \ \ \ \ \ \ \ \ \ \ \ \ \ \ \ }}{}$$ 
\vspace{-8mm}
$$\leftrightarrow$$
\vspace{-7mm}
$$(i \text{ has inductive reason to believe } A) \land (A \text{ indicates to $i$ that } B)$$
\noindent Now observe for any proposition $W$ and true evidence $E$ that $W \land E$ entails $W$. Hence, $W$ always indicates to $i$ that $W$. Similarly if $P$ is valid then for any proposition $W$ and true evidence $E$ we have that $W \land E$ entails $P$, i.e. $W$ indicates to $i$ that $P$. Thus, we have as additional primitive rules:
$$\infer{A \text{ indicates to } i \text{ that } A}{} \ \ \ \ \ \infer{A \text{ indicates to } i \text{ that } B}{B}$$
Next suppose $W$ indicates to $i$ that $P$ and $W$ also indicates to $i$ that $P \to Q$. If $E$ is some true evidence which gives $i$ inductive reason to believe $W$ then $W \land E$ entails both $P$ and $P \to Q$, i.e. $W \land E$ entails $Q$. Hence, we also have that $W$ indicates to $i$ that $Q$. This gives us the primitive rule:
$$\infer
{[(A \text{ indicates to } i \text{ that } B) \land (A \text{ indicates to } i \text{ that } B \to C)] \to (A \text{ indicates to } i \text{ that } C)}{}$$ 
The above rule is quite powerful. For instance, along with our first three primitive rules, it can be used to show all the following derived rules:
$$\infer
{(i \text{ has $A$ as deductive reason to believe } B) \land [i \text{ has $A$ as deductive reason to believe } (B \to C)]}{}$$ 
\vspace{-8mm}
$$\to$$
\vspace{-7mm}
$$i \text{ has $A$ as deductive reason to believe } C$$

$$\infer
{[(A \text{ indicates to } i \text{ that } B) \land (A \text{ indicates to } i \text{ that } C)] \to (A \text{ indicates to } i \text{ that } B \land C)}{}$$

$$\infer
{(i \text{ has $A$ as deductive reason to believe } B) \land (i \text{ has $A$ as deductive reason to believe } C)}{}$$ 
\vspace{-7mm}
$$\to$$
\vspace{-7mm}
$$i \text{ has $A$ as deductive reason to believe } (B \land C)$$

$$\infer{A \text{ indicates to } i \text{ that } B}{A \to B}$$

$$\infer{(A \text{ indicates to } i \text{ that } B) \to (A \text{ indicates to } i \text{ that } C)}{B \to C}$$

$$\infer{(i \text{ has } A \text{ as deductive reason to  believe } B) \to (i \text{ has } A \text{ as deductive reason to  believe } C)}{B \to C}$$
Additionally, we will maintain throughout that our agents are \emph{valid reasoners}, i.e. if $W$ holds and $i$ has $W$ as deductive reason to believe $P$ then $P$ also holds. 
\newpage
\noindent To see why, note whenever $i$ has $W$ as deductive reason to believe $P$, there is some true evidence $E$ which gives $i$ inductive reason to believe $W$. Further, $W \land E$ entails $P$. As a result, if $W$ holds then $W \land E$ holds and so $P$ must hold as well. Thus, we have the rule:
$$\infer
{[A \land (i \text{ has } A \text{ as deductive reason to believe } B)] \to B}{}$$
However, we do not take the above rule to be primitive. For, next, we introduce into our syntax the phrase `$i$ has \emph{some true reason} to believe $P$.' Namely, we say $i$ has some true reason to believe $P$ iff some `witness' $W$ holds such that $i$ has $W$ as deductive reason to believe $P$. Then, to ensure our agents are valid reasoners, we require that $P$ holds whenever $i$ has some true reason to believe $P$. More compactly, this corresponds to the primitive rules: 
$$\infer
{[A \land (i \text{ has $A$ as deductive reason to believe } B)] \to (i \text{ has some true reason to believe } B)}{}$$ 

$$\infer
{(i \text{ has some true reason to believe } A) \to A}{}$$
\noindent We will at times alternatively gloss `$i$ has some true reason to believe $P$' instead as `$i$ \emph{can inductively know} $P$.' For recall that if $i$ has inductive reason to believe some $W$ then $i$ can feasibly and efficiently limit decide $W$ after affirming it in light of some true evidence. Hence, if such a $W$ holds, $i$ can attain stable justified true belief in $W$. If $W$ further indicates to $i$ that $P$ (so $i$ has $W$ as deductive reason to believe $P$) then $i$ can also attain stable justified true belief in $P$ by choosing to believe $P$ whenever they choose to believe $W$. Therefore, $i$ can inductively know $P$ whenever $i$ has some true reason to believe $P$. \footnote[3]{As an example, suppose Alice has agreed to deposit assets of a particular value in a designated account as collateral to Bob and her account initially starts with sufficient collateral. At every future time step, Bob can see the assets in Alice's account and determine whether their total value continues to constitute a sufficient amount of collateral. Unfortunately, the proposition $P=$`Alice's account eventually always has sufficient collateral' is not limit decidable for Bob. However, for any finite natural $k$, the proposition $W_k=$`Alice's account dips below having sufficient collateral at most $k$ times but eventually always has sufficient collateral' is limit decidable for Bob. In particular, if  Bob has a switching tolerance of at least $2k+1$, then Bob can feasibly and efficiently limit decide $W_k$ after affirming it in light of initially seeing that Alice's account has sufficient collateral, i.e. Bob has inductive reason to believe $W_k$. Suppose in reality that Alice's account dips below having sufficient collateral exactly $1$ time before bouncing back to eventually always having sufficient collateral and Bob has a switching tolerance of $5$. Then $W_2$ is true and so Bob can attain stable justified true belief in the proposition `Alice's account dips below having sufficient collateral at most $2$ times but eventually always has sufficient collateral.' Additionally, since $W_2$ entails $P$, Bob has $W_2$ as deductive reason to believe $P$ and can therefore also attain stable justified true belief in $P$. In this sense, Bob can inductively know `Alice's account eventually always has sufficient collateral' even though he cannot limit decide this fact.}
\\
\\
Our secondary thesis will be that the following infinite list of conditions \textbf{L1} constitute what it means for a witness $W$ to generate common inductive knowledge of a proposition $P$:
\begin{enumerate}
\item Everyone has inductive reason to believe $W$.
\item $W$ indicates to everyone that $P$.
\item $W$ indicates to everyone that $1$ and $2$.
\item[] ... \ \ \ \ \ ... \ \ \ \ \ ... \ \ \ \ \ ... \ \ \ \ \ ... \ \ \ \ \ ... \ \ \ \ \ ... \ \ \ \ \ ...
\item[$n$.] $W$ indicates to everyone that $1$ and $(n-1)$.
\end{enumerate}
Using 1 and 2, we obtain the primitive rule:
$$\infer
{(A \text{ generates common inductive knowledge of } B) \to (i \text{ has } A \text{ as deductive reason to believe } B)}{}$$ 
Now, note $1$ and $3$ entail `Everyone has $W$ as deductive reason to believe $1$ and $2$.' Since $1$ and $2$ entail `Everyone has $W$ as deductive reason to believe $P$,' $1$ and $3$ also entail `Everyone has $W$ as deductive reason to believe `everyone has $W$ as deductive reason to believe $P$.'' Similarly, starting with $1$ and $4$, one can recursively derive `Everyone has $W$ as deductive reason to believe `everyone has $W$ as deductive reason to believe `everyone has $W$ as deductive reason to believe $P$.''' Hence, \textbf{L1} entails what we will call \textbf{L2}:
\begin{enumerate}
\item[$(1)$] Everyone has $W$ as deductive reason to believe $P$.
\item[$(2)$] Everyone has $W$ as deductive reason to believe $(1)$.
\item[$(3)$] Everyone has $W$ as deductive reason to believe $(2)$.
\item[] ... \ \ \ \ \ ... \ \ \ \ \ ... \ \ \ \ \ ... \ \ \ \ \ ... \ \ \ \ \ ... \ \ \ \ \ ... \ \ \ \ \ ...
\item[$(n)$] Everyone has $W$ as deductive reason to believe $(n-1)$.	
\end{enumerate}
Next, note $(2)$ entails `$W$ indicates to everyone that $(1)$.' Since $(1)$ entails 1 and 2, $(2)$ also entails `$W$ indicates to everyone that 1 and 2,' i.e. $(2)$ entails 1 and 3. Similarly, starting with $(3)$, one can recursively derive `$W$ indicates to everyone that 1 and 3,' i.e. $(3)$ entails 1 and 4. Thus, \textbf{L1} and \textbf{L2} are logically equivalent and so either can be taken to constitute what it means for $W$ to generate common inductive knowledge of $P$. While \textbf{L1} evidently bears more structural similarity to Lewis' account of how common inductive knowledge is generated, we will work frequently with \textbf{L2} for purposes of clarity. For instance, \textbf{L2} makes it obvious that if $W$ generates common inductive knowledge of $P$ then for each $(n) \in \textbf{L2}$ everyone has $W$ as deductive reason to believe $(n)$, and so given our prior pre-theoretic commitments, everyone has $W$ as deductive reason to believe `$W$ generates common inductive knowledge of $P$.' This motivates the primitive rule of inference:
$$\infer
{\text{ \ \ \ \ \ \ \ \ \ \ \ \ \ \ \ \ \ \ \ \ \ \ \ \ \ \ \ \ \ \ \ \ } A \text{ generates common inductive knowledge of } B \text{ \ \ \ \ \ \ \ \ \ \ \ \ \ \ \ \ \ \ \ \ \ \ \ \ \ \ \ \ \ \ \ \ }}{}$$ 
\vspace{-9.5mm}
$$\to$$
\vspace{-8mm}
$$i \text{ has $A$ as deductive reason to believe } (A \text{ generates common inductive knowledge of } B)$$
\noindent In fact, for any proposition $Q$, it can be seen that if $Q$ entails `everyone has $W$ as deductive reason to believe $P$' and $Q$ entails `everyone has $W$ as deductive reason to believe $Q$' then $Q$ entails every proposition in $\textbf{L2}$ holds. To demonstrate, using the first entailment and the consequent of the second entailment, we have $Q$ entails `everyone has $W$ as deductive reason to believe `everyone has $W$ as deductive reason to believe $P$.'' Similarly, using this new entailment with the consequent of the second entailment, one can recursively derive $Q$ entails `everyone has $W$ as deductive reason to believe `everyone has $W$ as deductive reason to believe `everyone has $W$ as deductive reason to believe $P$.'''  Thus, we stipulate as yet another primitive rule:
$$C \to (\text{Everyone has } A \text{ as deductive reason to believe } C) $$
$$\infer{C \to (A \text{ generates common inductive knowledge of } B)}{C \to (\text{Everyone has } A \text{ as deductive reason to believe } B)}$$
How then ought we define common inductive knowledge? Lewis suggests we say $P$ can become common inductive knowledge iff :
\begin{quote}
\centering
some state of affairs holds such that it generates common inductive knowledge of $P$.\footnote[4]{Again, strictly speaking, Lewis actually says that $P$ is common knowledge iff some state of affairs holds such that it is a basis for common knowledge of $P$.}
\end{quote}
While we agree with Lewis that this condition is sufficient for common inductive knowledge, we will insist that it is not necessary.
\newpage
\noindent To motivate why, suppose that $W$ holds and generates common inductive knowledge of $P$ as Lewis suggests. Then, using \textbf{L2}, we immediately have \textbf{L3}:
\begin{enumerate}
\item[$(1)'$] Everyone has some true reason to believe $P$.
\item[$(2)'$] Everyone has some true reason to believe $(1)'$.
\item[$(3)'$] Everyone has some true reason to believe $(2)'$.
\item[] ... \ \ \ \ \ ... \ \ \ \ \ ... \ \ \ \ \ ... \ \ \ \ \ ... \ \ \ \ \ ... \ \ \ \ \ ... \ \ \ \ \ ...
\item[$(n)'$] Everyone has some true reason to believe $(n-1)'$.	
\end{enumerate}
Recall Lewis seeks to show his 3 conditions for a witness to generate common inductive knowledge of $P$ entail exactly this sort of infinite list. But \textbf{L3} is not equivalent to Lewis' definition of common inductive knowledge for each agent $i$ could have a different true reason for believing each proposition in \textbf{L3}. On our view, Lewis' definition is unnecessarily stringent by his own lights and \textbf{L3} is closer to a desirable definition of common inductive knowledge. However, \textbf{L3} is not without its own shortcomings.  For if every proposition in \textbf{L3} holds, although we can conclude for each $(n)' \in \textbf{L3}$ everyone has some true reason to believe $(n)'$, we cannot further conclude everyone has some true reason to believe every $(n)' \in \textbf{L3}$. After all, agents may not have a principled procedure for creating a new `combined' true reason to believe every $(n)' \in \textbf{L3}$. Thus, to guarantee our notion of common inductive knowledge is a `fixed point,' our tertiary thesis will be that what it means for $P$ to potentially become common inductive knowledge is constituted by satisfying the primitive rules of inference:
$$\infer
{(A \text{ can become common inductive knowledge}) \to A}{}$$ 

$$\infer
{\text{ \ \ \ \ \ \ \ \ \ \ \ \ \ \ \ \ \ \ \ \ \ \ \ \ \ \ \ \ } A \text{ can become common inductive knowledge} \text{ \ \ \ \ \ \ \ \ \ \ \ \ \ \ \ \ \ \ \ \ \ \ \ \ \ \ \ \ }}{}$$ 
\vspace{-7.5mm}
$$\to$$
\vspace{-7mm}
$$i \text{ has some true reason to believe } (A \text{ can become common inductive knowledge})$$

$$\infer
{A \to(B \text{ can become common inductive knowledge})}{A \to (\text{Everyone has some true reason to believe } A) \text{ \ \ \ } A \to B}$$
In particular, since `everyone has some true reason to believe' can be alternatively glossed as `everyone can inductively know,' this means `$P$ can become common inductive knowledge' is the weakest proposition entailing $P$ such that if it is true then everyone can inductively know it. 
\\
\\
As a check, we now briefly argue why Lewis' definition of common inductive knowledge entails ours. So, suppose $P$ can become Lewisian common inductive knowledge. Then some `witness' $W$ holds so that $W$ generates common inductive knowledge of $P$.  By our earlier inference rules, this means everyone has $W$ as deductive reason to believe $P$ and also that everyone has $W$ as deductive reason to believe `$W$ generates common inductive knowledge of $P$.' Since $W$ holds and everyone is a valid reasoner, it follows that $P$ holds and so `$P$ can become Lewisian common inductive knowledge' entails $P$. Additionally, since $W$ entails $W$, it follows that everyone has $W$ as deductive reason to believe `$W$ holds and generates common inductive knowledge of $P$.' Using the fact $W$ holds then, this means `$P$ can become Lewisian common inductive knowledge' entails `everyone has some true reason to believe `$P$ can become Lewisian common inductive knowledge.''  Therefore, by our last primitive rule, we can conclude `$P$ can become Lewisian common inductive knowledge' entails `$P$ can become common inductive knowledge,' i.e.:
$$\infer
{[W \land (W \text{ generates common inductive knowledge of } P)]}{}$$
\vspace{-8.5mm}
$$\to$$
\vspace{-7.5mm}
$$P \text{ can become common inductive knowledge}$$
 
\newpage
\noindent In the coming section, we provide the reader with an overview of concepts from topological learning theory which will feature prominently in our semantics. We begin by reviewing standard results from the field. This review is intended to fix notation and recall basic facts. Readers who are already familiar with the topological difference hierarchy and its epistemic interpretation may wish to skip ahead. Such readers can turn directly to page 12 where we begin introducing several new pieces of terminology. These new terms and the theorems we prove about them will be essential for understanding the rest of the paper.
  
\section{Topology \& Learning}
Let $\Omega \neq \varnothing$ be the set of all possible worlds. We say that a \emph{piece of evidence} $E$ for agent $i$ is a non-empty subset of  possible worlds in $\Omega$ which eliminates all worlds outside $E$ as possibilities. Let $\mathcal{E}_i$ consist of all pieces of evidence agent $i$ might ever learn. Further, given a world $w \in \Omega$, let $\mathcal{E}_{i}(w)$ denote those elements of $\mathcal{E}_i$ which contain $w$. Intuitively, $\mathcal{E}_{i}(w)$ represents exactly those pieces of evidence which agent $i$ eventually learns in world $w$. We will assume that $\mathcal{E}_i$ forms an \emph{information basis} for $i$, that is:
\begin{enumerate}
\item $\forall w \in \Omega, \mathcal{E}_{i}(w) \neq \varnothing$.
\item $\forall w \in \Omega$ and $\forall E_1, E_2 \in \mathcal{E}_{i}(w)$ there exists $E_3 \in \mathcal{E}_{i}(w)$ such that $E_3 \subseteq E_1 \cap E_2$.
\end{enumerate}
The first requirement above simply says that agent $i$ learns some piece of evidence in every world. In our view, this is a very weak assumption. For even if agent $i$ never rules out any possible world in $\Omega$, then $\Omega$ itself must be a piece of evidence for $i$. The second requirement is more substantive and corresponds to the assumption that evidence is cumulative. Namely, this requirement says that if agent $i$ learns both $E_1$ and $E_2$ as evidence then they will also learn some piece of evidence $E_3$ at least as strong as $E_1 \cap E_2$. We now note that our definition of an \emph{information} basis is identical to that of a \emph{topological} basis. Hence, the collection $\mathcal{T}_i$ generated by taking arbitrary unions of pieces of evidence in $\mathcal{E}_i$ forms a topology over $\Omega$. 
\\
\\
In particular, it turns out the limit decidability of a set for agent $i$ will be closely related to whether the set can be expressed as the nested difference of a certain number of descending open sets in the topology $\mathcal{T}_i$.\footnote[5]{We define the nested difference of a finite sequence of sets $S_0$,...,$S_n$ to be given by $S_0 \backslash (S_1 \backslash (... S_n)...)$.} However, to illustrate why, we will need to introduce quite a bit of terminology. For starters, say that:
\begin{itemize}
\item A \emph{decision method} for $i$ is a map $m_i : \mathcal{E}_i \to \{\text{Yes}, \text{No}\}$.
\item A $t$\emph{-switching sequence starting with `Yes'} for $m_i$ is a finite downward sequence $E_0 \supseteq ... \supseteq E_t$ of pieces of evidence in $\mathcal{E}_i$ such that $m_i(E_{2k})=\text{Yes}$ and $m_i(E_{2k+1})=\text{No}$.
\item A $t$\emph{-switching sequence starting with `No'} for $m_i$ is a finite downward sequence $E_0 \supseteq ... \supseteq E_t$ of pieces of evidence in $\mathcal{E}_i$ such that $m_i(E_{2k})=\text{No}$ and $m_i(E_{2k+1})=\text{Yes}$.
\item A decision method for $i$ \emph{has at most $n$ switches after saying `Yes'} if it is a decision method which contains no $t$-switching sequences starting with `Yes' for any $t>n$.
\item A decision method for $i$ \emph{has at most $n$ switches after saying `No'} if it is a decision method which contains no $t$-switching sequences starting with `No' for any $t>n$.
\item A decision method for $i$ \emph{has a bounded number of switches} if it is a decision method which either has at most $n$ switches after saying `Yes' or has at most $n$ switches after saying `No' for some $n$.
\end{itemize}
\textbf{Lemma 3.0:} Given a decision method $m_i$ for agent $i$ which has a bounded number of switches, there exists a function $\sigma_{m_i}: \Omega \to \{\text{Yes}, \text{No}\}$ which $\forall w \in \Omega$ sets:
$$\sigma_{m_i}(w)=\begin{cases}\text{Yes} & \exists E \in \mathcal{E}_{i}(w) \forall E' \in \mathcal{E}_{i}(w), \ E' \subseteq E \to m_i(E')=\text{Yes} \\ \text{No} & \exists E \in \mathcal{E}_{i}(w) \forall E' \in \mathcal{E}_{i}(w), \ E' \subseteq E \to m_i(E')=\text{No} \end{cases}$$ 
\underline{Proof} 
\\
First we show $\sigma_{m_i}$ is a well-defined partial function. So suppose towards a contradiction it were not. Then for some $w^* \in \Omega$ both piecewise conditions above are satisfied. Then there $\exists E_1,E_2 \in \mathcal{E}_{i}(w^*)$ such that $\forall E' \in \mathcal{E}_{i}(w^*), \ E' \subseteq E_1 \to m_i(E')=\text{Yes}$ and  $\ E' \subseteq E_2 \to m_i(E')=\text{No}$. Further, since $\mathcal{E}_i$ is an information basis, $\exists E_3 \in \mathcal{E}_{i}(w^*)$ such that $E_3 \subseteq E_1 \cap E_2$. But this implies that both $m_i(E_3)=\text{Yes}$ and $m_i(E_3)=\text{No}$, giving us our desired contradiction. 
\\
Now we show $\sigma_{m_i}$ is total. So suppose towards a contradiction it were not. Then for some $w^* \in \Omega$ neither piecewise condition above is satisfied. Further, since $\mathcal{E}_i$ is an information basis, we know $\exists E_0 \in \mathcal{E}_{i}(w^*)$. WLOG suppose $m_i(E_0)=\text{Yes}$. Because the first piecewise is not satisfied, $\exists E_1 \in \mathcal{E}_{i}(w^*)$ so that $E_1 \subseteq E_0$ and $m_i(E_1)=\text{No}$. However, because the second piecewise condition is not satisfied, $\exists E_2 \in \mathcal{E}_{i}(w^*)$ so that $E_2 \subseteq E_1$ and $m_i(E_2)=\text{Yes}$. In this way we can create an infinite downward sequence $E_0 \supseteq E_{1} \supseteq E_{2} ...$ of pieces of evidence in $\mathcal{E}_i$ such that $m_i(E_{2k})=\text{Yes}$ and $m_i(E_{2k+1})=\text{No}$. This contradicts the fact that $m_i$ is a decision method which has a bounded number of switches, as desired. $\square$
\\
\\
Conceptually speaking, given a decision method $m_i$ for agent $i$ which has a bounded number of switches, $\sigma_{m_i}(w)$ returns the output that $m_i$ converges to in world $w$. Since $\sigma^{-1}_{m_i}(\text{Yes})=W$ iff $\sigma^{-1}_{m_i}(\text{No})=\Omega \backslash W$:
\begin{itemize}
	\item Agent $i$ \emph{can limit decide $W$ in $n$ switches after saying `Yes'} just in case there is a decision method $m_i$ for $i$ such that $m_i$ has at most $n$ switches after saying `Yes' and $\sigma^{-1}_{m_i}(\text{Yes})=W$.
	\item Agent $i$ \emph{can limit decide $W$ in $n$ switches after saying `No'} just in case there is a decision method $m_i$ for $i$ such that $m_i$ has at most $n$ switches after saying `No' and $\sigma^{-1}_{m_i}(\text{Yes})=W$.
\end{itemize}
\noindent \textbf{Theorem 3.0:} Agent $i$ can limit decide $W$ in $n$ switches after saying `Yes' if and only if $W$ can be expressed as the nested difference of $n+1$ descending open sets in $\mathcal{T}_i$.
\\
\underline{Proof} 
\\
First we prove the `only if' direction. So assume agent $i$ can limit decide $W$ in $n$ switches after saying `Yes.' Then there is a decision method $m_i$ for $i$ such that $m_i$ has at most $n$ switches after saying `Yes'  and $\sigma^{-1}_{m_i}(\text{Yes})=W$. Let $I_0=\{E \in \mathcal{E}_i :  m_i(E)=\text{Yes} \}$. Next, recursively define $I_k$ by setting:
$$I_k=\begin{cases} \{E' \in \mathcal{E}_i : \exists E \in I_{k-1}, E' \subseteq E \text{ and }  m_i(E')=\text{No}\} & k \text{ is odd} \\ \{E' \in \mathcal{E}_i : \exists E \in I_{k-1}, E' \subseteq E \text{ and }  m_i(E')=\text{Yes}\} & k \text{ is even} \end{cases}$$ 
Note that $I_{n+1}=\varnothing$ since if $\exists E_{n+1} \in I_{n+1}$ then we could find a ($n+1$)-switching sequence starting with `Yes' for $m_i$  of the form $E_0 \supseteq ... \supseteq E_{n+1}$ where each $E_k \in I_k$. Now let $O_k= \bigcup_{E \in I_k} E$. Each $O_k$ is open as it is a union of basis elements. In particular, $\{O_k\}_{k \in \mathbb{N}}$ forms a descending sequence of open sets with $O_{n+1}=\varnothing$ and so it suffices show $O_0 \backslash (O_1 \backslash (... O_n)...)=W$. 
\\
Suppose $w \in O_0 \backslash (O_1 \backslash (... O_n)...)$. Let $k^*$ be the largest $k$ for which $w \in O_k$. Since $O_{k^*}$ is not empty, we know $k^*<n+1$. Additionally, since $w \in O_0 \backslash (O_1 \backslash (... O_n)...)$, we know $k^*$ is even. Hence, $\exists E^* \in I_{k^*}$ such that $w \in E^*$ and $m_i(E^*)=\text{Yes}$. Assume for sake of contradiction that $\exists E' \in \mathcal{E}_{i}(w)$ such that $E' \subseteq E^*$ and $m_i(E')=\text{No}$. Then $E' \in I_{k^*+1}$ and so $w \in E' \subseteq O_{k^*+1}$, contradicting maximality of $k^*$. Thus, $E^*$ is an element of $\mathcal{E}_{i}(w)$ such that $\forall E' \in \mathcal{E}_{i}(w)$ we have  $E' \subseteq E^* \to m_i(E')=\text{Yes}$, i.e. $w \in \sigma^{-1}_{m_i}(\text{Yes})=W$. 
\newpage
\noindent Now suppose $w \in W=\sigma^{-1}_{m_i}(\text{Yes})$. Then $\exists E_0 \in \mathcal{E}_{i}(w)$ such that $m_i(E_0)=\text{Yes}$. In particular, $E_0 \in I_0$ and so $w \in E_0 \subseteq O_0$. Let $k^*$ be the largest $k$ for which $w \in O_k$. Since $O_{k^*}$ is not empty we know $k^* < n+1$. Suppose towards a contradiction that $k^*$ is odd. Then, $\exists E^* \in I_{k^*}$ such that $w \in E^*$ and $m_i(E^*)=\text{No}$. Further $E^*$ is an element of $\mathcal{E}_{i}(w)$ such that $\forall E' \in \mathcal{E}_{i}(w)$, if $E' \subseteq E^*$ then $m_i(E')=\text{No}$ (otherwise we would have $E' \in I _{k^*+1}$ and $w \in E' \subseteq O_{k^*+1}$, contradicting maximality of $k^*$). But this means $w \in \sigma^{-1}_{m_i}(\text{No})$. Contradiction. Thus $k^*$ is even and so we can conclude $w \in O_0 \backslash (O_1 \backslash (... O_n)...)$.
\\
\\
Next we prove the `if' direction. So suppose $W$ can be expressed as the nested difference of $n+1$ descending open sets $O_0 \supseteq ... \supseteq O_n$ in $\mathcal{T}_i$.  Given $E\in \mathcal{E}_i$, define $k_E$ to be the largest $k$ for which $E \subseteq O_k$ (if there is no such $k$ set $k_E=-1$). Let $m_i$ be the decision method for $i$ defined by $\forall E \in \mathcal{E}_i$ setting:
$$m_i(E)=\begin{cases} \text{No} & k_E \text{ is odd}\\ \text{Yes} & k_E \text{ is even}\end{cases}$$
Assume for sake of contradiction there is a ($n+1$)-switching sequence starting with `Yes' for $m_i$ of the form $E_0 \supseteq ... \supseteq E_{n+1}$. Since $k_{E_0}$ is at least $0$ and $k_{E_0}<k_{E_1}<...<k_{E_{n+1}}$, we know $k_{E_{n+1}} \geq n+1$. However, we also know  $k_{E_{n+1}} \leq n$ by definition. Contradiction. Hence, $m_i$ has at most $n$ switches after saying `Yes' and so it suffices to show $O_0 \backslash (O_1 \backslash (... O_n)...)=\sigma_{m_i}^{-1}(\text{Yes})$. 
\\
Suppose $w \in O_0 \backslash (O_1 \backslash (... O_n)...)$. Let $k^*$ be the largest $k$ for which $w \in O_k$. Then $\exists E^* \in \mathcal{E}_{i}(w)$ such that $E^* \subseteq O_{k^*}$. Further, $E^*$ is an element $\mathcal{E}_{i}(w)$ such that $\forall E' \in \mathcal{E}_{i}(w)$, if $E' \subseteq E^*$ then $k_{E'}=k^*$ (otherwise we would have $k_{E'}>k^*$ and $w \in E' \subseteq O_{k_{E'}}$, contradicting maximality of $k^*$). Since $w \in O_0 \backslash (O_1 \backslash (... O_n)...)$, we know $k^*$ is even. Thus $\forall E' \in \mathcal{E}_{i}(w)$, if $E' \subseteq E^*$ then $m_i(E')=\text{Yes}$, i.e. $w \in \sigma_{m_i}^{-1}(\text{Yes})$. 
\\
Now suppose $w \in \sigma^{-1}_{m_i}(\text{Yes})$.  Then $\exists E_0 \in \mathcal{E}_{i}(w)$ such that $m_i(E_0)=\text{Yes}$. In particular, $w \in E_0 \subseteq O_0$. Let $k^*$ be the largest $k$ for which $w \in O_k$. Suppose for contradiction that $k^*$ is odd. We know $\exists E^* \in \mathcal{E}_{i}(w)$ such that $E^* \subseteq O_{k^*}$. Further, $E^*$ is an element $\mathcal{E}_{i}(w)$ such that $\forall E' \in \mathcal{E}_{i}(w)$, if $E' \subseteq E^*$ then $k_{E'}=k^*$ (otherwise we would have $k_{E'}>k^*$ and $w \in E' \subseteq O_{k_{E'}}$, contradicting maximality of $k^*$). But since $k^*$ is odd this means $\forall E' \in \mathcal{E}_{i}(w)$, if $E' \subseteq E^*$ then $m_i(E')=\text{No}$, i.e. $w \in \sigma_{m_i}^{-1}(\text{No})$. Contradiction. Thus $k^*$ is even and so we can conclude $w \in O_0 \backslash (O_1 \backslash (... O_n)...)$. $\square$
\\
\\
\noindent \textbf{Lemma 3.1:} Agent $i$ can limit decide $W$ in $n$ switches after saying `No' if and only if agent $i$ can limit decide $\Omega \backslash W$ in $n$ switches after saying `Yes.'
\\
\underline{Proof} 
\\
We only show one direction as the other direction follows from a parallel argument. So assume agent $i$ can limit decide $W$ in $n$ switches after saying `No.' Then there is a decision method $m_i$ for $i$ such that $m_i$ has at most $n$ switches after saying `No' and $\sigma^{-1}_{m_i}(\text{Yes})=W$. Consider the decision method $m'_i$ obtained by $\forall E \in \mathcal{E}_i$ setting:
$$m'_i(E)=\begin{cases}\text{Yes} & m_i(E)=\text{No} \\ \text{No} & m_i(E)=\text{Yes}\end{cases}$$ 
Note that $m'_i$ has at most $n$ switches after saying `Yes.' Further $\sigma_{m'_i}^{-1}(\text{Yes})=\sigma_{m_i}^{-1}(\text{No})=\Omega \backslash W$. Thus agent $i$ can limit decide $\Omega \backslash W$ in $n$ switches after saying `Yes.' $\square$
\\
\\
\textbf{Corollary 3.0:} Agent $i$ can limit decide $W$ in $n$ switches after saying `No' if and only if $W$ can be expressed as the complement of a nested difference of $n+1$ descending open sets in $\mathcal{T}_i$.
\\
\underline{Proof} 
\\
Immediate by the Theorem 3.0 and Lemma 3.1. $\square$
\newpage
\noindent Going forward, we will say a set is \emph{$n$-open} in $\mathcal{T}_i$ iff it is the nested difference of $n$ descending open sets in $\mathcal{T}_i$. Additionally, say a set is \emph{$n$-closed} in $\mathcal{T}_i$ iff it is the complement of a set which is $n$-open in $\mathcal{T}_i$. Finally, say a set is \emph{$n$-clopen} in $\mathcal{T}_i$ iff it is both $n$-open and $n$-closed in $\mathcal{T}_i$. These topological complexity classes are then ordered by inclusion according to the below \emph{topological difference hierarchy}:\footnote[6]{For historical context, Putnam's 1965 paper was the first to try and characterize the complexity of what he called `trial and error predicates,' albeit in a computability theoretic context. Kevin Kelly later recast the problem in general topological terms...}
\vspace{-3mm}
\begin{figure}[h]
\centering
\includegraphics[width=12cm]{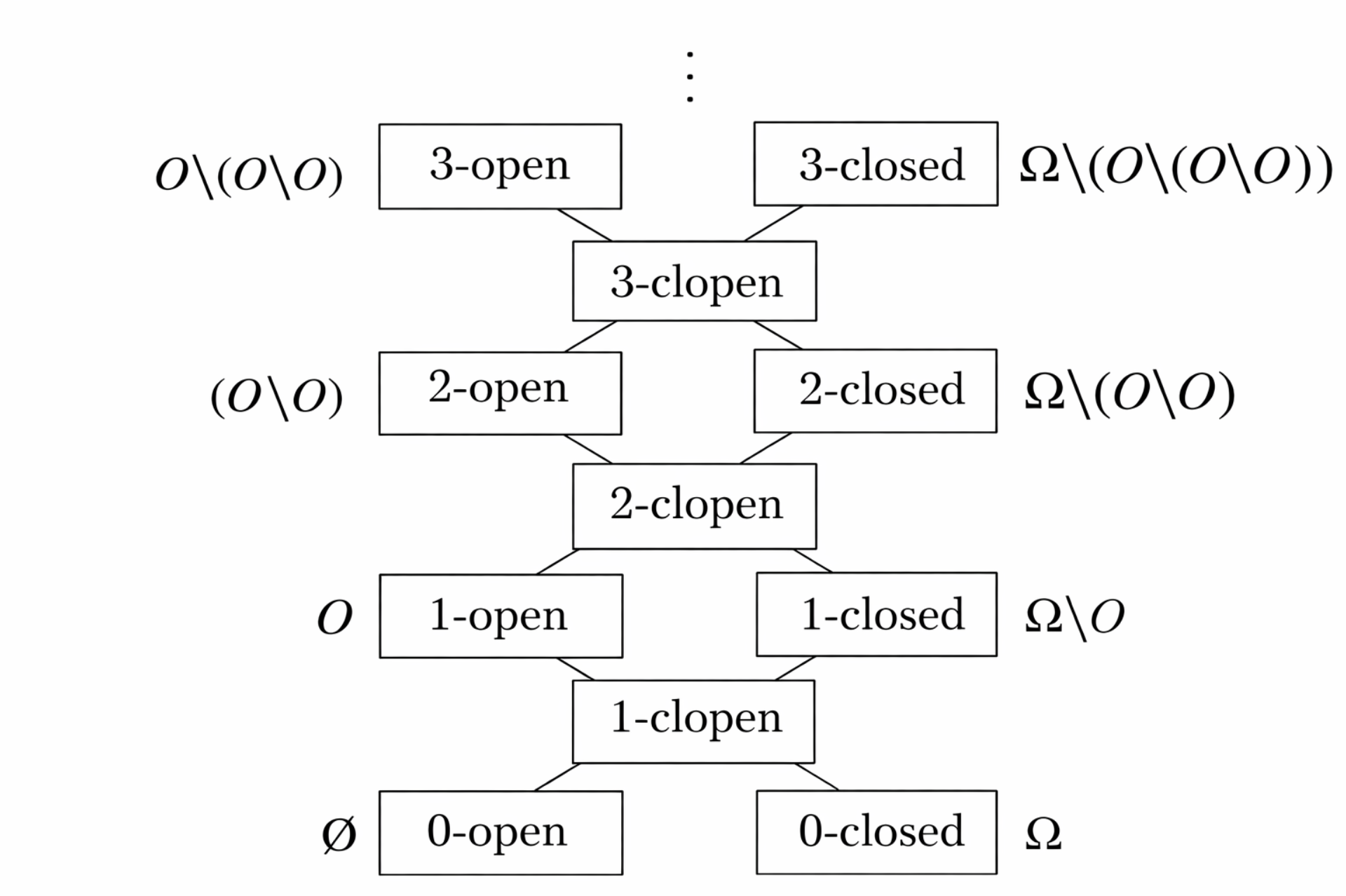}
\end{figure}
\\
We have already seen an epistemic interpretation for when $n>0$ of what it means to be $n$-open, $n$-closed, and $n$-clopen in $\mathcal{T}_i$. But what does it mean for a set to be 0-open or 0-closed in $\mathcal{T}_i$?\footnote[7]{No set is $n$-open, $n$-closed, or $n$-clopen for $n<0$.} One interpretation is as follows: say an information basis $\mathcal{E}_i$ for $i$ \emph{has a starting point} if $\Omega \in \mathcal{E}_i$, i.e. agent $i$ receives $\Omega$ as a piece of evidence before learning anything stronger. If $\mathcal{E}_i$ has a starting point, a decision method $m_i$ for $i$ must make a choice of whether $m_i(\Omega)=\text{No}$ or $m_i(\Omega)=\text{Yes}$. Intuitively then, a set $W$ is $0$-open in $\mathcal{T}_i$ if and only if agent $i$ can limit decide $W$ in $0$ switches \emph{starting from} `No,' i.e. $W=\varnothing$. Similarly, a set $W$ is $0$-closed in $\mathcal{T}_i$ if and only if agent $i$ can limit decide $W$ in $0$ switches \emph{starting from} `Yes,' i.e. $W=\Omega$.
\\
\\
We now temporarily assume agent $i$'s information basis $\mathcal{E}_i$ has a starting point and formalize what it means for agent $i$  to be able to limit decide $W$ in $n$ switches starting from `Yes'/`No' by saying:
\begin{itemize}
	\item A decision method $m_i$ for $i$ \emph{starts from `Yes' (`No')} if $m_i(\Omega)=\text{Yes}$ (No).
	\item A decision method for $i$ \emph{has at most $n$ switches starting from `Yes' (`No')} if it is a decision method which starts from `Yes' (`No') and has at most $n$ switches after saying `Yes' (`No').
	\item A decision method for $i$ \emph{has at most $n$ switches} if it is a decision method which has at most $n$ switches starting from `Yes' or at most $n$ switches starting from `No.'	
	\item Agent $i$ \emph{can limit decide $W$ in $n$ switches starting from `Yes' (`No')} if there is a decision method $m_i$ for $i$ such that $m_i$ has at most $n$ switches starting from `Yes' (`No') and $\sigma^{-1}_{m_i}(\text{Yes})=W$.
	\item Agent $i$ \emph{can limit decide $W$ in $n$ switches} if agent $i$ can limit decide $W$ in $n$ switches starting from `Yes' or in $n$ switches starting from `No.'
\end{itemize}
The above notions are also closely related to the topological difference hierarchy. To see precisely how, given a set $W$, let $n_{i,W}$ be the smallest $n$ for which agent $i$ can limit decide $W$ in $n$ switches (if there is no such $n$, let $n_{i,W}=\infty$). Additionally, assume agent $i$ has finite \emph{switching tolerance} $n_i \geq 0$ so that $i$ adopts only those decision methods which have at most $n_i$ switches. Because $i$ can `feasibly' limit decide $W$ iff  $n_{i,W} \leq n_i$ and does so `efficiently' iff their method has at most $n_{i,W}$ switches:
\begin{itemize}
\item Agent $i$ \emph{has reason to limit decide $W$ starting from `Yes' (`No')} just in case $n_{i,W} \leq n_i$ and agent $i$ can limit decide $W$ in $n_{i,W}$ switches starting from `Yes' (`No').
\end{itemize}
\textbf{Theorem 3.1:} Agent $i$ has reason to limit decide $W$ starting from `Yes' if and only if $\exists k \leq n_i$ such that $W$ is $k$-closed in $\mathcal{T}_i$ but not $(k-1)$-open.
\\
\underline{Proof} 
\\
First we show the `if' direction. 
\\
Suppose $\exists k \leq n_i$ so that $W$ is $k$-closed in $\mathcal{T}_i$ but not $(k-1)$-open. 
\\
If $k=0$ we know $W$ is $0$-closed, and hence, $i$ can limit decide $W$ in $0$ switches starting from `Yes,' i.e. $n_{i,W}=0$.  Since $0=n_{i,W} \leq n_i$, we can conclude agent $i$ has reason to limit decide $W$ starting from `Yes.' 
\\
If $k=1$ we know $W$ is not $0$-open, and hence, $i$ cannot limit decide $W$ in $0$ switches starting from `No.' However, because $W$ is $1$-closed, $i$ can limit decide $W$ in $0$ switches after saying `No.' Thus, there is a method $m_i$ for $i$ such that $m_i$ has at most $0$ switches after saying `No' and $\sigma^{-1}_{m_i}(\text{Yes})=W$. Further, we know $m_i(\Omega) \neq \text{`No'}$ as otherwise $i$ could limit decide $W$ in $0$ switches starting from `No.' Hence $m_i(\Omega) = \text{`Yes'}$ and so $i$ can limit decide $W$ in $1$ switch starting from `Yes.'  Therefore, either $n_{i,W}=0$ or $n_{i,W}=1$. Further $1=k \leq n_i$ and so $n_{i,W} \leq n_i$. If $n_{i,W}=1$  then we can conclude agent $i$ has reason to limit decide $W$ starting from `Yes.' Similarly, if $n_{i,W}=0$ then $i$ can limit decide $W$ in $0$ switches starting from `Yes' (as $i$ can't limit decide $W$ in $0$ switches starting from `No'). So, in this case as well, we can conclude agent $i$ has reason to limit decide $W$ starting from `Yes.' 
\\
If $k \geq 2$ we know $W$ is not $(k-1)$-open, and hence, $i$ cannot limit decide $W$ in $k-2$ switches after saying `Yes.' However because $W$ is $k$-closed, $i$ can limit decide $W$ in $k-1$ switches after saying `No.' Thus, there is a method $m_i$ for $i$ such that $m_i$ has at most $k-1$ switches after saying `No' and $\sigma^{-1}_{m_i}(\text{Yes})=W$. Further, for any such $m_i$, we know $m_i(\Omega) \neq \text{`No'}$ as otherwise $i$ could limit decide $W$ in $k-2$ switches after saying `Yes.' Hence $m_i(\Omega)=\text{Yes}$ and so $i$ can limit decide $W$ in $k$ switches starting from `Yes' but cannot in $(k-1)$ switches starting from `No.' Therefore, $n_{i,W} \leq k \leq n_i$. If $n_{i,W}=k$ then we can conclude agent $i$ has reason to limit decide $W$ starting from `Yes.' Similarly, if $n_{i,W}<k$ then $i$ can limit decide $W$ in $n_{i,W}$ switches starting from `Yes' (as $i$ can't limit decide $W$ in $n_{i,W}$ switches starting from `No'). So in these cases as well, we can conclude $i$ has reason to limit decide $W$ starting from `Yes.'
\\
\\
Now we show the `only if' direction. 
\\
Suppose agent $i$ has reason to limit decide $W$ starting from `Yes.' Then $n_{i,W} \leq n_i$. Further, $i$ can limit decide $W$ in $n_{i,W} $ switches starting from `Yes' and, hence, after saying `Yes'. Thus, $W$ is $n_{i,W}$-closed in $\mathcal{T}_i$ and so it suffices to show $W$ is not $(n_{i,W}-1)$-open in $\mathcal{T}_i$.
\\
If $n_{i,W}=0$ then, since $W$ is never $(-1)$-open, we are done.
\\
If $n_{i,W}=1$ then, by definition, $i$ cannot limit decide $W$ in $0$ switches. In particular, $i$ cannot limit decide $W$ in $0$ switches starting from `No.' Hence $W \neq \varnothing$ and so $W$ is not $0$-open as desired.
\\
If $n_{i,W} \geq 2$ then, by definition, $i$ cannot limit decide $W$ in $(n_{i,W}-1)$ switches. Suppose towards a contradiction $W$ is $(n_{i,W}-1)$-open. Then $i$ can limit decide $W$ in $n_{i,W}-2$ switches after saying `Yes.' Thus, there is a method $m_i$ for $i$ such that $m_i$ has at most $n_{i,W}-2$ switches after saying `Yes' and $\sigma^{-1}_{m_i}(\text{Yes})=W$. Further, we know $m_i(\Omega) \neq \text{Yes}$ as otherwise $i$ could limit decide $W$ in $(n_{i,W}-2)$ switches starting from `Yes.' Hence $m_i(\Omega) = \text{No}$. But this means agent $i$ can limit decide $W$ in $(n_{i,W}-1)$ switches starting from `No.' Contradiction. Hence $W$ is not $(n_{i,W}-1)$-open, as desired. $\square$
\newpage
\noindent \textbf{Lemma 3.2:} Agent $i$ has reason to limit decide $W$ starting from `No' if and only if agent $i$ has reason to limit decide $\Omega \backslash W$ starting from 	`Yes.'
\\
\underline{Proof} 
\\
We only show one direction as the other direction follows from a parallel argument. So assume agent $i$ has reason to limit decide $W$ starting from `No.' Then $n_{i,W} \leq n_i$ and there is a decision method $m_i$ for $i$ such that $m_i$ has at most $n_{i,W}$ switches starting from `No' and $\sigma^{-1}_{m_i}(\text{Yes})=W$. Consider the decision $m'_i$ obtained by $\forall E \in \mathcal{E}_i$ setting:
$$m'_i(E)=\begin{cases}\text{Yes} & m_i(E)=\text{No} \\ \text{No} & m_i(E)=\text{Yes}\end{cases}$$ 
Note that $m'_i$ has at most $n_{i,W}$ switches starting from `Yes.' Further $\sigma_{m'_i}^{-1}(\text{Yes})=\sigma_{m_i}^{-1}(\text{No})=\Omega \backslash W$. Hence $n_{i, \Omega \backslash W} \leq n_{i,W} \leq n_i$. Suppose towards a contradiction  $n_{i, \Omega \backslash W}<n_{i,W}$. Then we can construct a decision method $m''_i$ for $i$ such that $m''_i$ has at most $n_{i, \Omega \backslash W}$ switches and $\sigma_{m''_i}^{-1}(\text{Yes})=\Omega \backslash W$.  Further, we can construct a decision method $m'''_i$ from $m''_i$ (analogously to how we constructed $m'_i$ from $m_i$) such that $m'''_i$ has at most $n_{i, \Omega \backslash W}$ switches and $\sigma_{m'''_i}^{-1}(\text{Yes})=W$. Contradiction. Hence $n_{i, \Omega \backslash W}=n_{i,W}$ and we can conclude agent $i$ has reason to limit decide $\Omega \backslash W$ starting from `Yes.' $\square$
\\
\\
\textbf{Corollary 3.1:} Agent $i$ has reason to limit decide $W$ starting from `No' if and only if $\exists k \leq n_i$ such that $W$ is $k$-open in $\mathcal{T}_i$ but not $(k-1)$-closed.
\\
\underline{Proof} 
\\
Immediate by Theorem 3.1 and Lemma 3.2. $\square$
\\
\\
We now no longer assume $\mathcal{E}_i$ has a starting point. Instead, fix $E \in \mathcal{E}_i$ and let $\mathcal{E}_i | E=\{E' \in \mathcal{E}_i: E'\subseteq E\}$. If, in some state of affairs, $E$ comprises the set of all worlds $i$ considers possible then $\mathcal{E}_i | E$ consists of all those pieces of evidence which refine $E$ that $i$ might learn going forward. In particular, $\mathcal{E}_i | E$ forms a basis for the subspace topology $\mathcal{T}_i | E$ over $E$ where $E \in \mathcal{E}_i | E$.  Therefore, we can consistently relativize every bit of terminology we have introduced thus far to $E$ by replacing $\Omega$ with $E$ as the set of all possible worlds and replacing $\mathcal{E}_i$ with $\mathcal{E}_i|E$ so that $i$'s information basis has a starting point. In this context, a decision method $m_i$ for $i$ is a map $m_i: \mathcal{E}_i |E \to \{\text{Yes},\text{No}\}$. Furthermore, we can show that agent $i$ has reason to limit decide $W$ starting from `Yes' iff $\exists k \leq n_i$ such that $W$ is $k$-closed in $\mathcal{T}_i | E$ but not $(k-1)$-open.  Similarly, we can show that $i$ has reason to limit decide $W$ starting from `No' iff $\exists k \leq n_i$ such that $W$ is $k$-open in $\mathcal{T}_i | E$ but not $(k-1)$-closed. Implicit in both these statements however is that $W$ is a subset of $E$. So, given an arbitrary subset $W$ of $\Omega$, one must take its intersection with $E$ for such statements to even make sense (after all $W \cap E$ consists of those worlds where $W$ is true and $i$ learns evidence $E$). As a consequence then, in our original context, we say:
\begin{itemize}
\item Agent $i$ \emph{has reason to limit decide $W$ starting from `Yes' in light of evidence $E$} if $\exists k \leq n_i$ such that $W \cap E$ is $k$-closed in $\mathcal{T}_i | E$ but not $(k-1)$-open.
\item Agent $i$ \emph{has reason to limit decide $W$ starting from `No' in light of evidence $E$} if $\exists k \leq n_i$ such that $W \cap E$ is $k$-open in $\mathcal{T}_i | E$ but not $(k-1)$-closed.
\end{itemize}
For concision, we abbreviate the above phrases in natural language by writing:
\begin{itemize}
\item $E$ \emph{gives} $i$ \emph{inductive reason to believe} $W$.
\item $E$ \emph{gives} $i$ \emph{inductive reason to disbelieve} $W$.
\end{itemize}
and denote them by the predicates $\text{Yes}_i(W|E)$ and $\text{No}_i(W|E)$ respectively. 
\newpage
\noindent \textbf{Lemma 3.3:} Suppose $(X_j)_{j \in J}$ is a collection of $(n+1)$-open sets in $\mathcal{T}_i$. Furthermore, suppose  that each $X_j$ is the nested difference of a descending sequence of open sets $O^j_0 \supseteq ... \supseteq O_n^j$. If for each $m\in J$ we have $X_m=O^m_0 \cap \bigcup_{j \in J} X_j$ then $\bigcup_{j \in J} X_j$ is the nested difference of the descending sequence of open sets $\bigcup_{j \in J} O^j_0 \supseteq ... \supseteq \bigcup_{j \in J} O^j_n$ and so $\bigcup_{j \in J} X_j$ is $(n+1)$-open in $\mathcal{T}_i$.\footnote[8]{In general, when $n>0$, the union of two $(n+1)$-open sets in $\mathcal{T}_i$ may not itself be $(n+1)$-open. For instance, suppose we are working in Cantor space. Take $A$ to be the set of sequences which have at least $1$ and at most $5$ non-$0$ entries. Take $B$ to be the set of sequences which have at least $7$ and at most $13$ non-$0$ entries. One can verify that while $A$ and $B$ are both $2$-open, $A \cup B$ is not $2$-open (although, it is $4$-open). Similarly, when $n>1$, the intersection of two $(n+1)$-open sets in $\mathcal{T}_i$ may not itself be $(n+1)$-open. Again, suppose we are working in Cantor space. Take $C$ to be the set of sequences which have at least $3$ and at most $9$ non-$0$ entries. Take $D$ to be the set of sequences which have at least $11$ and at most $15$ non-$0$ entries. Defining $A$ and $B$ as before, one can verify that while $A \cup B$ and $C \cup D$ are both $4$-open, $(A \cup B) \cap (C \cup D)$ is not $4$-open (although, it is $6$-open).} 
\\
\underline{Proof} 
\\
We proceed by induction on $n$. When $n=0$, each $X_j=O^j_0$ and so we can immediately establish the base case as $\bigcup_{j \in J} X_j=\bigcup_{j \in J} O^j_0$. For the inductive step, assume the claim holds when $n=k-1$. Take $(X_j)_{j \in J}$ to be a collection of $(k+1)$-open sets in $\mathcal{T}_i$ so that each $X_j$ is the nested difference of a descending sequence of open sets $O^j_0 \supseteq .... \supseteq O_k^j$. Suppose for each $m\in J$ we have $X_m=O^m_0 \cap \bigcup_{j \in J} X_j$. Next, for each $j \in J$, let $X'_j$ be the nested difference of the descending sequence of open sets $O^j_1 \supseteq ... \supseteq O_k^j$.  We claim now that:
$$\bigcup_{j \in J} X_j=\left(\bigcup_{j \in J} O^j_0\right) \backslash \left(\bigcup_{j \in J} X'_j\right).$$ 
Suppose $w \in \left(\bigcup_{j \in J} O^j_0\right) \backslash \left(\bigcup_{j \in J} X'_j\right)$ then there is some $s\in J$ such that $w \in O^s_0$. Since we are guaranteed $w \notin X'_s$ this means $w \in X_s$ and so $w \in \bigcup_{j \in J} X_j$. Now suppose $w \in \bigcup_{j \in J} X_j$. Then there is some $s \in J$ such that $w \in X_s$. In particular, we know $w \in O^s_0$. Entertain for contradiction there exists some $m \in J$ so that $w\in X'_m$. Since $X_m=O^m_0 \cap \bigcup_{j \in J} X_j$  we know $O^m_0 \cap X_s \subseteq X_m$. Since $ w\in X'_m$ we have $w \notin X_m$  and hence $w \notin O^m_0 \cap X_s $. However, since $X'_m \subseteq O^m_0$, this is only possible if $w \notin X_s$. Contradiction. Therefore $w \notin \bigcup_{j \in J} X'_j$ and so $w \in \left(\bigcup_{j \in J} O^j_0\right) \backslash \left(\bigcup_{j \in J} X'_j\right)$. Next we claim for each $m\in J$ that:
$$X'_m=O^m_0 \cap \left(\bigcup_{j \in J} X_j\right)^c.$$ 
Suppose $w \in O^m_0 \cap \left(\bigcup_{j \in J} X_j\right)^c$. Then $w \in O^m_0$ and $w \notin \bigcup_{j \in J} X_j$.  Since $X_m=O^m_0 \cap \bigcup_{j \in J} X_j$, we know $w \notin X_m$. Further, because $ w\in O^m_0$, this is only possible if $w \in X'_m$. Now suppose $w \in X'_m$. Then $w \in O^m_0$ and $w \notin X_m$. Since $X_m=O^m_0 \cap \bigcup_{j \in J} X_j$, this is only possible if $w \notin \bigcup_{j \in J} X_j$. Hence, $w \in O^m_0 \cap \left(\bigcup_{j \in J} X_j\right)^c$. Now, using the fact $X'_m \subseteq O^m_1 $ and $X'_m \subseteq O^m_0 \subseteq \bigcup_{j \in J} O^j_0$, we have for each $m\in J$ that:
$$X'_m=\left(O^m_0 \cap \left(\bigcup_{j \in J} X_j\right)^c\right) \cap \left(O^m_1 \cap \bigcup_{j \in J} O^j_0\right).$$ 
Since $O^m_1 \subseteq O^m_0$ and $\left(\bigcup_{j \in J} X_j\right)^c \cap \left(\bigcup_{j \in J} O^j_0\right)=\bigcup_{j \in J} \left(O^j_0 \cap \left(\bigcup_{j' \in J} X_{j'}\right)^c\right)=\bigcup_{j \in J} X'_j$, we can rearrange the above to obtain for each $m \in J$ that $X'_m=O^m_1 \cap \bigcup_{j \in J} X'_j.$ 
By hypothesis then, we have $\bigcup_{j \in J} X'_j$ is the nested difference of the descending sequence of open sets $\bigcup_{j \in J} O^j_1 \supseteq ... \supseteq \bigcup_{j \in J} O^j_k$. Finally, since $\bigcup_{j \in J} X_j=\left(\bigcup_{j \in J} O^j_0\right) \backslash \left(\bigcup_{j \in J} X'_j\right)$, we can conclude $\bigcup_{j \in J} X_j$ is the nested difference of the descending sequence of open sets $\bigcup_{j \in J} O^j_0 \supseteq ... \supseteq \bigcup_{j \in J} O^j_k$. $\square$
\newpage
\noindent \textbf{Lemma 3.4:} If $W$ is $(n+1)$-open in $\mathcal{T}_i$ then $\forall w \in W$ there $\exists E \in \mathcal{E}_i(w)$ such that $W \cap E$ is $n$-closed in the subspace topology $\mathcal{T}_i|E$.
\\
\underline{Proof}
\\
Since $W$ is $(n+1)$-open there exists a descending sequence of open sets $O_0 \supseteq ... \supseteq O_{n}$ in $\mathcal{T}_i$ such that $W=O_0 \backslash (O_1 \backslash (... O_{n})...)$. Suppose $w \in W$. Then $w \in O_0$ and so there exists $E^* \in \mathcal{E}_i(w)$ such that $E^* \subseteq O_0$. Additionally:
$$W\cap E^*=(O_0 \cap E^*) \backslash ((O_1 \cap E^*) \backslash (... (O_{n} \cap E^*))...)=E^* \backslash ((O_1 \cap E^*) \backslash (... (O_{n} \cap E^*))...)$$
, i.e. $W \cap E^*$ is $n$-closed in $\mathcal{T}_i |E^*$. $\square$
\\
\\
\textbf{Theorem 3.2:} $W$ is $(n_i+1)$-open in $\mathcal{T}_i$ iff $\forall w \in W$ there $\exists E \in \mathcal{E}_i(w)$ such that $\text{Yes}_i(W |E)$ holds.
\\
\underline{Proof}
\\
First we show the `if' direction. Suppose $\forall w \in W$ there $\exists E \in \mathcal{E}_i(w)$ such that $\text{Yes}_i(W |E)$ holds. In particular, for each $w \in W$, take $E_w$ to be an element of $\mathcal{E}_i(w)$ for which $\text{Yes}_i(W |E_w)$ holds so that $W=\bigcup_{w \in W} W \cap E_w$. Note for each $w \in W$ there is a decision method $m_{i,w}: \mathcal{E}_i | E_w \to \{\text{Yes},\text{No}\}$ such that $m_{i,w}$ has at most $n_i$ switches starting from `Yes' and $\sigma^{-1}_{m_{i,w}}(\text{Yes})=W \cap E_w$. Therefore, $W\cap E_w$ is the nested difference of $n_i+1$ descending open sets $O^w_0 \supseteq ... O^w_{n_i}$ in $\mathcal{T}_i|E_w$ with $O^w_0=E_w$ (as per the construction in Theorem 3.0). Thus, for each $w' \in W$, we have $W \cap E_{w'}=O^{w'}_0 \cap W=O^{w'}_0  \cap \bigcup_{w \in W} (W \cap E_{w})$. Applying Lemma 3.3 then, we can conclude $W=\bigcup_{w \in W} W \cap E_w$ is $(n_i+1)$-open in $\mathcal{T}_i$.  
\\
\\
Now we show the `only if' direction. Suppose $W$ is $(n_i+1)$-open and $w \in W$. By Lemma 3.4, we can define $k_w \leq n_i$ to be the least $k$ for which $\exists E \in \mathcal{E}_{i}(w)$ so that $W \cap E$ is $k$-closed in $\mathcal{T}_i | E$ and let $E_w \in \mathcal{E}_{i}(w)$ be an information state for which $W \cap E_w$ is $k_w$-closed in $\mathcal{T}_i | E_w$. 
\\
If $k_w=0$ then $W \cap E_w$ is $0$-closed in $\mathcal{T}_i|E_w$ but not $(-1)$-open and so $\text{Yes}_i(W | E_w)$ holds. 
\\
If $k_w=1$ then $W \cap E_w$ is $1$-closed in $\mathcal{T}_i | E_w$ but not $0$-open (as $w\in W \cap E_w$) and so $\text{Yes}_i(W | E_w)$ holds. 
\\
If $k_w \geq 2$ then $W \cap E_w$ is $k_w$-closed in $\mathcal{T}_i | E_w$. Suppose for contradiction that $W \cap E_w$ is $(k_w-1)$-open in $\mathcal{T}_i|E_w$. Then there is a decision method $m_i: \mathcal{E}_i | E_w \to \{\text{Yes},\text{No\}}$  such that $m_i$ has at most $k_w-2$ switches after saying `Yes' and $\sigma^{-1}_{m_i}(\text{Yes})=W \cap E_w$. Because $w \in W \cap E_w$, there $\exists E' \in (\mathcal{E}_i | E_w)(w)$ for which $m_i(E')=\text{Yes}$. Consider the decision method $m'_i:\mathcal{E}_i| E' \to \{\text{Yes},\text{No\}}$ which $\forall E \in \mathcal{E}_i | E'$ sets $m'_i(E)=m_i(E)$. Clearly, $m'_i$ has at most $k_w-2$ switches starting from `Yes.'  Consider arbitrary $w' \in E'$. If $\sigma_{m_i}(w')=\text{Yes}$ then $\exists E'' \in (\mathcal{E}_i|E_w)(w')$ so that $\forall E \in (\mathcal{E}_i|E'')(w')$ we have $m_i(E)=\text{Yes}$. In particular, if we take $E''' \in (\mathcal{E}_i|E')(w')$ to be a subset of $E' \cap E''$ containing $w'$, then this means $\forall E \in (\mathcal{E}_i|E''')(w')$ we have $m'_i(E)=m_i(E)=\text{Yes}$. Hence, $\sigma_{m'_i}(w')=\text{Yes}$. Symmetrically, if $\sigma_{m_i}(w')=\text{No}$ then $\sigma_{m'_i}(w')=\text{No}$. Thus, it follows $\forall w' \in E'$ that $\sigma_{m'_i}(w')=\sigma_{m_i}(w')$. Namely, $\sigma^{-1}_{m'_i}(\text{Yes})=W \cap E'$ and so $W \cap E'$ is the nested difference of $k_w-1$ descending open sets $O_0 \supseteq...\supseteq O_{k_w-2}$ in $\mathcal{T}_i|E'$ with $O_0=E'$ (as per the construction in Theorem 3.0), i.e. $W \cap E'$ is $(k_w-2)$-closed in $\mathcal{T}_i|E'$. However, this contradicts minimality of $k_w$.  Therefore, $W \cap E_w$ is $k_w$-closed in $\mathcal{T}_i|E_w$ but not $(k_w-1)$-open and so $\text{Yes}_i(W | E_w)$ holds.
$\square$
\\
\\
\textbf{Corollary 3.2:} $W$ is $(n_i+1)$-closed in $\mathcal{T}_i$ iff $\forall w \in \Omega \backslash W$ there $\exists E \in \mathcal{E}_{i}(w)$ such that $\text{No}_i(W|E)$ holds.
\\
\underline{Proof}
\\
Immediate by Theorem 3.2 and the fact $\text{No}_i(W|E)$ holds iff $\text{Yes}_i(\Omega \backslash W|E)$ holds. $\square$   
\\
\\
In the next section, we will semantically interpret each of the pre-theoretic phrases we introduced earlier. For now, this concludes our overview of topological learning theory. We refer the curious reader to Kelly 1996 for a more comprehensive treatment of the subject as well as its relationship to computability.
\section{Semantics}
Define a frame $\mathcal{F}$ to be a pair $(\Omega,(\mathcal{E}_i,n_i)_{i \in N})$  where $\Omega \neq \varnothing$  and for each agent $i \in N$ we have that $(\mathcal{E}_i,n_i)$ consists of $i$'s information basis $\mathcal{E}_i$ over $\Omega$ together with $i$'s switching tolerance $n_i$. 
\\
\\
Next, given a frame $\mathcal{F}=(\Omega,(\mathcal{E}_i,n_i)_{i \in N})$, for each $i \in N$ define the operator $\textbf{R}_i: 2^{\Omega} \to 2^{\Omega}$ so that $\forall W \subseteq \Omega$:
$$\textbf{R}_i(W)=\{w \in \Omega : \exists E \in \mathcal{E}_i(w), \text{Yes}_i(W |E)\}$$
, i.e. $\textbf{R}_i(W)$ corresponds to those worlds where $i$ has inductive reason to believe $W$.\footnote[9]{Note we can use $\textbf{R}_i$ to now state Theorem 3.2 more concisely. Namely, we have that $W$ is $(n_i+1)$-open in $\mathcal{T}_i$ iff it is a postfixed point of the map $X \mapsto \textbf{R}_i(X)$.} 
\\
\\
Additionally, mirroring our discussion in the pre-theory section, given $W \subseteq \Omega$ define the operators $\textbf{I}_{i @ W}: 2^{\Omega} \to 2^{\Omega}$ and $\textbf{B}_{i @ W}: 2^{\Omega} \to 2^{\Omega}$ by $\forall P \subseteq \Omega$ setting:
$$\textbf{I}_{i@W}(P)=\{w \in \Omega : \forall E \in \mathcal{E}_{i}(w), \text{Yes}_i(W | E) \to W \cap E \subseteq P\} \text{ and }\textbf{B}_{i@W}(P)=\textbf{R}_i(W) \cap \textbf{I}_{i@W}(P)$$
, i.e. $\textbf{I}_{i@W}(P)$ and $\textbf{B}_{i@W}(P)$ correspond to the worlds where $W$ indicates to $i$ that $P$ and $i$ has $W$ as deductive reason to believe $P$ respectively.
\\
\\
Now, define $\textbf{S}_i: 2^{\Omega} \to 2^{\Omega}$ so that $\forall P \subseteq \Omega$:
$$\textbf{S}_i(P)=\{w \in \Omega: \exists W \subseteq \Omega, w \in W \cap \textbf{B}_{i @ W}(P)\}$$
, i.e. $\textbf{S}_i(P)$ corresponds to those worlds where $i$ has some true reason to believe $P$/can inductively know $P$.
\\
\\
Further, $\forall k \in \mathbb{N}^+$, let $\textbf{E}_{W}^k(P)$ denote the result of applying the map $X \mapsto \bigcap_{i \in N} \textbf{B}_{i@W}(X)$ iteratively $k$ times on $P$. Then, given $W \subseteq \Omega$, define $\textbf{G}_W: 2^{\Omega} \to 2^{\Omega}$ by $\forall P \subseteq \Omega$ setting:
$$\textbf{G}_W(P)=\bigcap_{k \in \mathbb{N^+}} \textbf{E}_{W}^k(P)$$ 
, i.e. $\textbf{G}_W(P)$ corresponds to those worlds where $W$ generates common inductive knowledge of $P$.
\\
\\
Finally, recall from the pre-theory section `$P$ can become common inductive knowledge' is supposed to be the weakest proposition entailing $P$ such that if it holds then everyone can inductively know it. However, this notion is well-defined relative to our frame $\mathcal{F}$ if and only if we can guarantee the existence of a greatest postfixed point for the map $X \mapsto P \cap \bigcap_{i \in N} \textbf{S}_i(X)$. We prove this is the case below.
\\
\\
\textbf{Lemma 4.0:} $\textbf{S}_i$ is monotone.
\\
\underline{Proof} 
\\
Suppose $P_1 \subseteq P_2$ and $w \in \textbf{S}_i(P_1)$. Then  for some $ W \subseteq \Omega$ there exists $E^* \in \mathcal{E}_{i}(w)$ such that $\text{Yes}_i(W|E)$. Further, $\forall E \in \mathcal{E}_{i}(w), \text{Yes}_i(W|E) \to W \cap E \subseteq P_1$. In particular, $\forall E \in \mathcal{E}_{i}(w), \text{Yes}_i(W|E) \to W \cap E \subseteq P_2$ as $P_1 \subseteq P_2$. Hence, $w \in \textbf{S}_i(P_2)$. $\square$
\\
\\
\textbf{Theorem 4.0:} The map $X \mapsto P \cap \bigcap_{i \in N}\textbf{S}_i(X)$ has a greatest (post)fixed point. 
\\
\underline{Proof}
\\
Since each $\textbf{S}_i$ is monotone, the map $X \mapsto P \cap \bigcap_{i \in N}\textbf{S}_i(X)$ is monotone and so Tarski's theorem guarantees the existence of a greatest fixed point which is equal to its greatest postfixed point. $\square$
\\
\newpage
\noindent But what can be said of this fixed point beyond its existence? In order to provide a more `constructive' characterization of what it means for $P$ to potentially become common inductive knowledge, we now study how each operator $\textbf{S}_i$ relates to the topology $\mathcal{T}_i$ generated by $\mathcal{E}_i$.
\\
\\
\textbf{Lemma 4.1:} Let $\text{int}_i$ be the interior operator with respect to the topology $\mathcal{T}_i$. If $n_i=0$ then $\forall P \subseteq \Omega$ we have that $\textbf{S}_i(P)=\text{int}_i(P)$.
\\\
\underline{Proof} 
\\
Firstly, note $\text{Yes}_i(W | E)$ holds for some piece of evidence $E \in \mathcal{E}_i$ iff $W \cap E$ is $0$-closed in $\mathcal{T}_i | E$, i.e. $E \subseteq W$. Now suppose $w \in \text{int}_i(P)$. Then $\exists W \in \mathcal{E}_{i}(w)$ such that $W \subseteq P$. Thus we immediately have that  $\forall E \in \mathcal{E}_{i}(w), \text{Yes}_i(W | E) \to W \cap E \subseteq P$. Further, since $w \in W$ and $\text{Yes}_i(W | W)$ holds, we can conclude $\exists E \in \mathcal{E}_{i}(w)$ such that $\text{Yes}_i(W |E)$ holds and therefore $w \in \textbf{S}_i(P)$. Next suppose $w \in \textbf{S}_i(P)$. Then $\exists W \subseteq \Omega$ containing $w$ for which $\exists E^* \in \mathcal{E}_{i}(w), E^* \subseteq W$ and $\forall E \in \mathcal{E}_{i}(w), E \subseteq W \to W \cap E \subseteq P$. In particular, this means $W \cap E^* \subseteq P$ and so $E^* \subseteq P$, i.e. $w \in \text{int}_i(P)$. $\square$
\\
\\
This leads us to ask whether, in general, $\textbf{S}_i$ acts like an `interior' operator with respect to some topology over $\Omega$. Towards that end, the Kuratowski interior axioms say a map $\textbf{i}: 2^{\Omega} \to 2^{\Omega}$ can be viewed as the interior operator with respect to some topology $\mathcal{T}_{\textbf{i}}$ over $\Omega$ iff:
\begin{enumerate}
\item \textbf{i} \emph{preserves the total space}, i.e. we have $\textbf{i}(\Omega)=\Omega$
\item \textbf{i} \emph{is intensive}, i.e. $\forall P \subseteq \Omega$ we have $\textbf{i}(P) \subseteq P$ 
\item \textbf{i} \emph{is idempotent}, i.e. $\forall P \subseteq \Omega$ we have $\textbf{i}(P)=\textbf{i}(\textbf{i}(P))$ 	
\item \textbf{i} \emph{preserves binary intersections}, i.e. $\forall P_1,P_2 \subseteq \Omega$ we have $\textbf{i}(P_1) \cap \textbf{i}(P_2)=\textbf{i}(P_1 \cap P_2)$ 
\end{enumerate}
Indeed, it turns out that $\textbf{S}_i$ satisfies all four of Kuratowski's axioms.
\\
\\
\textbf{Theorem 4.1:} $\textbf{S}_i$ preserves the total space.
\\
\underline{Proof} 
\\
It suffices to show $\Omega \subseteq \textbf{S}_i (\Omega)$. Suppose $w \in \Omega$. Since $\mathcal{E}_{i}(w)$ is non-empty, $\exists E^* \in \mathcal{E}_{i}(w)$. In particular, $\Omega \cap E^*=E^*$ is $0$-closed in $\mathcal{T}_i | E^*$ and so $\text{Yes}_i(\Omega | E^*)$ holds. Further $\forall E \in \mathcal{E}_{i}(w), \text{Yes}_i(\Omega | E) \to \Omega \cap E \subseteq \Omega$. Since $\Omega \subseteq \Omega$, we can conclude $w \in \textbf{S}_i (\Omega)$. $\square$
\\
\\
\textbf{Theorem 4.2:} $\textbf{S}_i$ is intensive.\footnote[10]{Putting together Theorems 4.1 and 4.2, we can transfinitely iterate the map $X \mapsto P \cap \bigcap_{i \in N}\textbf{S}_i(X)$ starting with $\Omega$ and obtain its greatest fixed point is the transfinite limit of the statements `Everyone has some true reason to believe $P$,' `Everyone has some true reason to believe `Everyone has some true reason to believe $P$.'' }
\\
\underline{Proof} 
\\
Suppose $w \in \textbf{S}_i(P)$. Then for some $W\subseteq \Omega$ containing $w$, we have that $\exists E^* \in\mathcal{E}_{i}(w)$ such that $\text{Yes}_i(W|E^*)$ holds. Further, $\forall E \in \mathcal{E}_{i}(w), \text{Yes}_i(W|E) \to W \cap E \subseteq P$. Thus $W \cap E^* \subseteq P$. Since $w \in E^*$, this means we can conclude $w \in P$. $\square$ 
\\
\\
\textbf{Theorem 4.3:} $\textbf{S}_i$ is idempotent.
\\
\underline{Proof} 
\\
Because $\textbf{S}_i$ is intensive we know $\textbf{S}_i(\textbf{S}_i(P)) \subseteq \textbf{S}_i(P)$ and so it suffices to show $\textbf{S}_i(P) \subseteq \textbf{S}_i(\textbf{S}_i(P))$. Suppose $w \in \textbf{S}_i(P)$. Then there exists $W \subseteq \Omega$ containing $w$  such that $\forall E \in \mathcal{E}_{i}(w), \text{Yes}_i(W|E) \to W \cap E \subseteq P$. Take arbitrary $E' \in \mathcal{E}_{i}(w)$ for which $\text{Yes}_i(W|E')$ holds. Then, $\exists k \leq n_i$ so that $W \cap E'$ is $k_i$-closed in $\mathcal{T}_{i}|E'$ but not $(k_i-1)$-open.  Hence, $\text{Yes}_i(W \cap E'|E')$ also holds. Take $w' \in W \cap E'$. Because $E' \in \mathcal{E}_{i}(w')$, we know $\exists E \in \mathcal{E}_{i}(w')$ for which $\text{Yes}_i(W \cap E'|E)$ holds. Further, $\forall E \in \mathcal{E}_{i}(w'), \text{Yes}_i(W \cap E' |E) \to (W \cap E') \cap E \subseteq P$ as $W \cap E' \subseteq P$. Hence, $w' \in \textbf{S}_i(P)$ and so $W \cap E' \subseteq \textbf{S}_i(P)$. Since $E'$ was arbitrary, this means that $\forall E \in \mathcal{E}_{i}(w), \text{Yes}_i(W|E) \to W \cap E \subseteq \textbf{S}_i(P)$. Also, because $w \in \textbf{S}_i(P)$, there exists $E^* \in \mathcal{E}_{i}(w)$ for which $\text{Yes}_i(W|E^*)$ holds. Therefore, we can conclude $w \in \textbf{S}_i(\textbf{S}_i(P))$. $\square$
\\
\\
By monotonicity of $\textbf{S}_i$, we already have $\forall P_1,P_2 \subseteq \Omega$ that $\textbf{S}_i(P_1 \cap P_2) \subseteq \textbf{S}_i(P_1)$ and $\textbf{S}_i(P_1 \cap P_2) \subseteq \textbf{S}_i(P_2)$, i.e. $\textbf{S}_i(P_1 \cap P_2) \subseteq \textbf{S}_i(P_1) \cap \textbf{S}_i(P_2)$. Hence, for $\textbf{S}_i$ to preserve intersections, it suffices to show $\forall P_1,P_2 \subseteq \Omega$ that $\textbf{S}_i(P_1) \cap \textbf{S}_i(P_2) \subseteq \textbf{S}_i(P_1 \cap P_2)$. Intuitively then, whenever $\textbf{S}_i(P_1)$ and $\textbf{S}_i(P_2)$ hold, we want a principled procedure which picks out some true reason $i$ has to believe $P_1$ as well as some true reason $i$ has to believe $P_2$ and combines them to produce some true reason $i$ has for believing $P_1 \cap P_2$. We now prove a few lemmas that will help construct exactly this procedure.
\\
\\
\textbf{Lemma 4.2:} $\textbf{S}_i (P)$ is the union of all $(n_i+1)$-open sets in $\mathcal{T}_i$ which entail $P$.
\\
\underline{Proof}
\\
Let $P(n_i)$ be the collection of all $(n_i+1)$-open sets in $\mathcal{T}_i$ which entail $P$. Suppose $w \in \bigcup_{W \in P(n_i)} W$. Then $\exists W \subseteq \Omega$ containing $w$ such that $W$ is $(n_i+1)$-open in $\mathcal{T}_i$ and $W \subseteq P$. Further, as $W \subseteq P$, we immediately have $\forall E \in \mathcal{E}_{i}(w), \text{Yes}_i(W | E) \to W \cap E \subseteq P$. In particular, by Theorem 3.2, $\exists E^* \in \mathcal{E}_{i}(w)$ for which $\text{Yes}_i(W | E)$ holds. Hence, $w \in \textbf{S}_i(P)$. Now suppose $w \in \textbf{S}_i(P)$. Then for some $W' \subseteq \Omega$ containing $w$ there exists $E^* \in \mathcal{E}_i(w)$ such that $\text{Yes}_i(W' | E^*)$ holds and $W '\cap E^* \subseteq P$. In particular, $W' \cap E^*$ is $n_i$-closed in $\mathcal{T}_i | E^*$ and so it is $(n_i+1)$-open in $\mathcal{T}_i$. Finally, because $w \in W' \cap E^*$, we have that $W' \cap E^* \in P(n_i)$ and so $w \in \bigcup_{W \in P(n_i)} W$. Thus, $\textbf{S}_i (P)=\bigcup_{W \in P(n_i)} W$ is the union of all $(n_i+1)$-open sets in $\mathcal{T}_i$ entailing $P$. $\square$
\\
\\
\textbf{Lemma 4.3:} $\forall n>0$, $P$ is a union of $(n+1)$-open sets in $\mathcal{T}_i$ iff $P$ is a union of $2$-open sets in $\mathcal{T}_i$.
\\
\underline{Proof} 
\\
Since the `if' direction is trivial, we focus on the `only if' direction. Suppose $P$ is a union of $(n+1)$-open sets in $\mathcal{T}_i$, i.e. there exists a collection $\mathcal{C}$ of $(n+1)$-open sets in $\mathcal{T}_i$ such that $P=\cup_{W \in \mathcal{C}} W$. Fix $W \in \mathcal{C}$. Since $W$ is $(n+1)$-open in $\mathcal{T}_i$ there exists a descending sequence of open sets $W_0 \supseteq ... \supseteq W_{n}$ in $\mathcal{T}_i$ such that $W=W_0 \backslash(W_1 \backslash( ... W_{n})...)$. For ease of argument, let $W_{n+1}=\varnothing$. Note $w \in W$ iff there is an even $k\neq n+1$ for which $w \in W_{k} \backslash W_{k+1}$, i.e. $W=\cup_{k \leq n: k \text{ is even}} W_k \backslash W_{k+1}$. Since each $W_k \backslash W_{k+1}$ is $2$-open in $\mathcal{T}_i$, it follows that $P$ is a union of $2$-open sets in $\mathcal{T}_i$. $\square$
\\
\\
\textbf{Lemma 4.4:} If $n_i>0$ then $\textbf{S}_i (P)$ is the union of all $2$-open sets in $\mathcal{T}_i$ which entail $P$.
\\
\underline{Proof}
\\
Let $P(2)$ be the collection of all $2$-open sets in $\mathcal{T}_i$ which entail $P$. Suppose $w \in \bigcup_{W \in P(2)} W$. Then $\exists W \subseteq \Omega$ containing $w$ such that $W$ is 2-open in $\mathcal{T}_i$ and $W \subseteq P$. Note $W$ is certainly $(n_i+1)$-open in $\mathcal{T}_i$ as it is 2-open. Hence, by Lemma 4.2, $w \in \textbf{S}_i(P)$. Now suppose $w \in \textbf{S}_i(P)$. Then, by Lemma 4.2, $w$ belongs to a set $W'$ which is $(n_i+1)$-open $\mathcal{T}_i$ and entails $P$. Additionally, by Lemma 4.3, $w$ belongs to a set $W$ which is $2$-open in $\mathcal{T}_i$ and entails $W'$. Finally, because $W \subseteq P$, we can conclude $w \in \bigcup_{W \in P(2)} W$. Thus, $\textbf{S}_i(P)=\bigcup_{W \in P(2)} W$ is the union of all 2-open sets in $\mathcal{T}_i$ entailing $P$. $\square$
\\
\\
The above implies that whenever $n_i>0$  if $w \in \textbf{S}_i(P)$ then $w$ is contained in a $2$-open set $W$ of $\mathcal{T}_i$ which entails $P$. Thus, if $i$ has some true reason to believe $P$ then $i$ also has some true $2$-open set as reason to believe $P$.  We will think of the principled procedure we are constructing as picking out such 2-open sets as reasons in case $n_i>0$. In the next lemma, we show that if one `combines' two 2-open reasons by taking their intersection then the resulting reason is also 2-open in $\mathcal{T}_i$.
\newpage
\noindent \textbf{Lemma 4.5:} If $W_1$ and $W_2$ are $2$-open in $\mathcal{T}_i$ then so is $W_1 \cap W_2$.
\\
\underline{Proof}
\\
Since $W_1$ and $W_2$ are $2$-open in $\mathcal{T}_i$, there exist descending sequences of open sets $O_0 \supseteq O_1$ and $O'_0 \supseteq O'_1$ such that $W_1=O_0 \backslash O_1$ and $W_2=O'_0 \backslash O'_1$. In particular:
$$W_1 \cap W_2=(O_0 \cap O'_0) \backslash (O_1 \cup O'_1)=(O_0 \cap O'_0) \backslash ((O_0 \cap O'_0) \cap (O_1 \cup O'_1))$$
Since $(O_0 \cap O'_0) \supseteq ((O_0 \cap O'_0) \cap (O_1 \cup O'_1))$, we can conclude $W_1 \cap W_2$ is $2$-open in $\mathcal{T}_i$.  $\square$
\\
\\
Conveniently then, using Lemmas 4.4 and 4.5, we can show...
\\
\\
\textbf{Theorem 4.4:} $\textbf{S}_i$ preserves binary intersections.
\\
\underline{Proof}
\\
By monotonicity, it suffices to show $\textbf{S}_i(P_1) \cap \textbf{S}_i(P_2) \subseteq \textbf{S}_i(P_1 \cap P_2)$. When $n_i=0$ we know $\textbf{S}_i$ is just $\text{int}_i$ and so we are done. Hence, suppose $n_i>0$ and $w \in \textbf{S}_i(P_1) \cap \textbf{S}_i(P_2)$. By Lemma 4.4, for some $W_1,W_2 \subseteq \Omega$ containing $w$ which are $2$-open in $\mathcal{T}_i$, we have that $W_1 \subseteq P_1$ and $W_2 \subseteq P_2$. By Lemma 4.5, $W_1 \cap W_2$ is $2$-open in $\mathcal{T}_i$. Thus $w$ is contained in a $2$-open set of $\mathcal{T}_i$ which entails $P_1 \cap P_2$, i.e. $w \in \textbf{S}_i(P_1 \cap P_2)$. $\square$
\\
\\
As promised then, by Theorems 4.1-4.4, $\textbf{S}_i$ satisfies the Kuratowski interior axioms. If agent $i$ is a deductive learner with $n_i=0$ then the topology $\mathcal{T}_{\textbf{S}_i}$ over $\Omega$ generated by $\textbf{S}_i$ is just $\mathcal{T}_i$. Alternatively, if agent $i$ is a truly inductive learner with $n_i>0$, Lemma 4.4 and Lemma 4.5 tell us $\mathcal{T}_{\textbf{S}_i}$ can be generated by taking the $2$-open sets of $\mathcal{T}_i$ as basis elements. In particular, so long as $n_i>0$, the topology $\mathcal{T}_{\textbf{S}_i}$ is invariant of $n_i$. Having detailed the properties of $\textbf{S}_i$, we now return to the question of how we can more constructively characterize the worlds where $P$ can become common inductive knowledge. 
\\
\\
\textbf{Theorem 4.5:} Let $\text{int}_N$ be the interior operator with respect to the intersection topology $\bigcap_{i \in N} \mathcal{T}_{\textbf{S}_i}$. The greatest fixed point of the map $X \mapsto P \cap \bigcap_{i \in N}\textbf{S}_i(X)$ is $\text{int}_N(P)$.
\\
\underline{Proof}
\\
Suppose $X$ is a fixed point of the map $X \mapsto P \cap \bigcap_{i \in N}\textbf{S}_i(X)$, i.e. $X = P \cap \bigcap_{i \in N}\textbf{S}_i(X)$. Clearly $X$ is a subset of $P$. Additionally for each $i \in N$, we have $X \subseteq \textbf{S}_i(X)$ and so $X= \textbf{S}_i(X)$ by intensivity. But this means $X$ is open in each $\mathcal{T}_{\textbf{S}_i}$, i.e. $X$ is open in $\bigcap_{i \in N} \mathcal{T}_{\textbf{S}_i}$. Hence, it follows that $X \subseteq \text{int}_N(P)$. Next, since $\text{int}_N(P)$ is a subset of $P$ and open in each $\mathcal{T}_{\textbf{S}_i}$, we know $\text{int}_N(P)= P \cap \bigcap_{i \in N}\textbf{S}_i(\text{int}_N(P))$. As a result, $\text{int}_N(P)$ is the greatest fixed point of the map $X \mapsto P \cap \bigcap_{i \in N}\textbf{S}_i(X)$. $\square$
\\
\\
Therefore, define the operator $\textbf{C}: 2^{\Omega} \to 2^{\Omega}$ so that $\forall P \subseteq \Omega$:
$$\textbf{C}(P)=\text{int}_N(P)$$
, i.e. $\textbf{C}(P)$ corresponds to those worlds where $P$ can become common inductive knowledge. Note if each $n_i>0$ then $\bigcap_{i \in N} \mathcal{T}_{\textbf{S}_i}$ is invariant of the $n_i$ and so $\textbf{C}(P)$ is as well. Thus, if agents are truly inductive learners, then the set of worlds where $P$ can become common inductive knowledge is invariant of agents' switching tolerances. We view this property of our semantics as a feature and not a bug. After all, a good notion of common inductive knowledge should be fairly robust to changes in agents' inductive standards. 
\\
\\
On the other hand, recall from the pre-theory section that Lewis says $P$ can become common inductive knowledge if and only if: 
\begin{quote}
\centering
some state of affairs holds such that it generates common inductive knowledge of $P$.
\end{quote}
Lewis' account lacks the robustness of our definition as, for any witnessing state of affairs which holds, whether that witness generates common inductive knowledge substantially varies with agents' switching tolerances. To illustrate, suppose $\Omega=\{w_1,w_2,w_3\}$ and take $P=\{w_1,w_3\}$. Further, suppose we have $|N|=2$ agents named Alice and Bob (who we abbreviate by writing $a$ and $b$ respectively). Finally, suppose Alice and Bob each have information bases over $\Omega$ given by the below diagrams:
\\
\vspace{-7.5mm}
\begin{figure}[h]
\centering
\includegraphics[width=9.5cm]{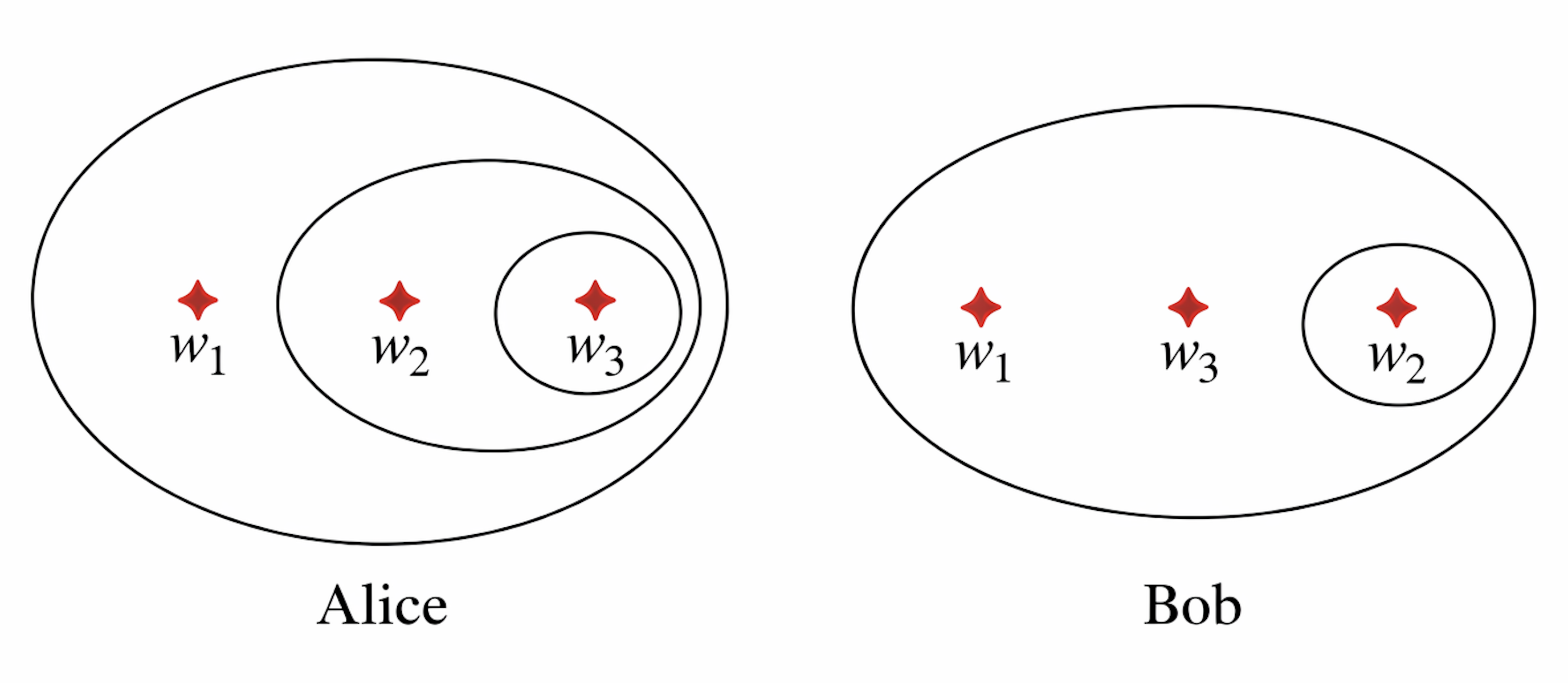}
\end{figure}
\vspace{-6mm}
\\
If $n_a=n_b=1$ it is easy to see the only witness $W$ that Bob has as deductive reason to believe $P$ in any world is $W=\{w_1,w_3\}$. One can also compute straightforwardly that $\textbf{B}_{a@W} (P)=\{w_3\}$. However, this means $\textbf{B}_{b@W}(\textbf{B}_{a@W} (P))=\varnothing$ and so $\textbf{G}_W(P)=\varnothing$. Hence, under Lewis' definition, $P$ cannot become common inductive knowledge in any world. Alternatively if $n_a=2$, one can compute that $\textbf{B}_{a@W} (P)=\textbf{B}_{b@W} (P)=\Omega$ and so $\textbf{G}_W(P)=\Omega$. Moreover, on Lewis' account, we suddenly have $P$ can become common inductive knowledge in every world $P$ holds as $W \cap \textbf{G}_W(P)=P$. In fact, one can generalize the above example so Lewis' definition shifts from saying $P$ cannot become common inductive knowledge in any world to saying $P$ can become common inductive knowledge in every world $P$ holds at arbitrarily high switching tolerances for Alice. By contrast, for all values of $n_a,n_b>0$, our definition of common inductive knowledge robustly maintains that $\textbf{C}(P)=P$. This can be seen visually using the topological characterization of $\textbf{C}$ in Theorem 4.5 almost immediately. For a more epistemic argument, note $W_1=\{w_1\}=\Omega \backslash \{w_2,w_3\}$ and $W_3=\{w_3\}$ are $2$-open sets in $\mathcal{T}_a$ which entail $P$. Hence, Alice has $W_1$ as true reason to believe $P$ at $w_1$ and $W_3$ as true reason to believe $P$ at $w_3$, i.e. $\textbf{S}_a(P)=P$. Similarly, $W_1 \cup W_3=\Omega \backslash \{w_2\}$ is a $2$-open set in $\mathcal{T}_b$ which entails $P$. Therefore, Bob has $W_1 \cup W_3$ as true reason to believe $P$ at both $w_1$ and $w_3$, i.e. $\textbf{S}_b(P)=P$. Thus, $P$ is the weakest proposition entailing $P$ such that if it is true then Alice and Bob can inductively know it, i.e. $\textbf{C}(P)=P$. 
\\
\\
In the sequel, we present the formal syntax for our logical language as well as a proof system with respect to which it is sound. Readers eager to see the relevance of our logic to distributed consensus should feel free to skip over to section 7. 
 
\section{Syntax}
Let $\text{PROP}$ be a countable set of primitive propositions and $N$ be a finite set of agents. Our logical language $\mathcal{L}$ will be defined by the Backus-Naur form:
$$\varphi ::= p | \top | \bot | \neg \varphi | \varphi_1 \land \varphi_2 | \varphi_1 \lor \varphi_2 | \varphi_1 \to \varphi_2 | \varphi_1 \leftrightarrow \varphi_2 | \mathbb{R}_i \varphi | \mathbb{I}_{i@\varphi_1} \varphi_2 | \mathbb{B}_{i@\varphi_1} \varphi_2| \mathbb{S}_i \varphi | \mathbb{G}_{\varphi_1} \varphi_2 | \mathbb{C} \varphi$$
where $p \in \text{PROP}$ and $i \in N$. 
\\
\\
The table on the following page summarizes how each of the modal connectives above correspond to each of the pre-theoretic phrases/semantic operators defined in the prior sections. 
\newpage
\begin{table}[h]
\centering
\begin{tabular}{|c|l|}
\hline
\textbf{Notation} & \multicolumn{1}{c|}{\textbf{Gloss}} \\
\hline
$\mathbb{R}_i \varphi$ & $i$ has inductive reason to believe $\varphi$ \\
$\mathbb{I}_{i@\varphi_1} \varphi_2$ & $\varphi_1$ indicates to $i$ that $\varphi_2$ \\
$\mathbb{B}_{i@\varphi_1} \varphi_2$ & $i$ has $\varphi_1$ as deductive reason to believe $\varphi_2$ \\
$\mathbb{S}_i \varphi$ & $i$ has some true reason to believe $\varphi /$can inductively know $\varphi$ \\
$\mathbb{G}_{\varphi_1} \varphi_2$ & $\varphi_1$ generates common inductive knowledge of $\varphi_2$ \\
$\mathbb{C} \varphi$ & $\varphi$ can become common inductive knowledge \\
\hline
\end{tabular}
\caption{Gloss of epistemic modalities}
\end{table}

\noindent More formally, a model $\mathcal{M}$ for $\mathcal{L}$ is a pair $(\mathcal{F},v)$ where $\mathcal{F}$ is a frame $(\Omega,(\mathcal{E}_i,n_i)_{i \in N})$ and $v: \text{PROP} \to 2^{\Omega}$ is a valuation function over primitive propositions. Then $\forall w\in \Omega$, given a primitive proposition $p \in \text{PROP}$ we write $(\mathcal{M},w) \vDash p \text{ iff } w \in v(p)$, i.e. $v(p)$ yields an interpretation of those worlds where $p$ holds in $\mathcal{M}$. Next, we extend $v$ so that its domain consists of all formulas $\varphi$ in $\mathcal{L}$ by interpreting classical connectives in their standard manner and recursively writing:
\begin{align*}
(\mathcal{M},w) &\vDash \mathbb{R}_i \varphi \text{ iff } w \in \textbf{R}_i(v(\varphi))	
\\
(\mathcal{M},w) &\vDash \mathbb{I}_{i@\varphi_1} \varphi_2 \text{ iff } w \in \textbf{I}_{i@ v(\varphi_1)} (v(\varphi_2))
\\
(\mathcal{M},w) &\vDash \mathbb{B}_{i@\varphi_1} \varphi_2 \text{ iff } w \in \textbf{B}_{i@ v(\varphi_1)}(v(\varphi_2))
\\
(\mathcal{M},w) &\vDash \mathbb{S}_{i} \varphi \text{ iff } w \in \textbf{S}_i(v(\varphi))
\\
(\mathcal{M},w) &\vDash \mathbb{G}_{\varphi_1} \varphi_2 \text{ iff } w \in \textbf{G}_{v(\varphi_1)}(v(\varphi_2))
\\
(\mathcal{M},w) &\vDash \mathbb{C} \varphi \text{ iff } w \in \textbf{C}(v(\varphi))
\end{align*}
, while $\forall \varphi \in \mathcal{L}$ setting $v(\varphi)=\{w \in \Omega: (\mathcal{M},w) \vDash \varphi\}$. 
\\
\\
Our proof system takes not only all propositional tautologies to be axioms, but also all instances of the following primitive schemas:
\[
\begin{array}{@{}l@{\qquad}l@{}}
\textbf{Ax}_{\mathsf{R}}:\ \mathbb{R}_i \varphi \leftrightarrow \mathbb{R}_i \mathbb{R}_i \varphi
&
\\[0.9em]

\textbf{Ax}_{\mathsf{I1}}:\ \mathbb{I}_{i@\varphi} \varphi
&
\textbf{Ax}_{\mathsf{I4}}:\ \mathbb{I}_{i@\varphi_1} \varphi_2 \rightarrow \mathbb{I}_{i@\varphi_1} \mathbb{R}_i(\varphi_1 \land \varphi_2)
\\

\textbf{Ax}_{\mathsf{I2}}:\ \mathbb{I}_{i@\varphi_1} \mathbb{I}_{i@\varphi_1} \varphi_2 \rightarrow \mathbb{I}_{i@\varphi_1} \varphi_2
&
\textbf{Ax}_{\mathsf{I5}}:\ (\mathbb{I}_{i@\varphi_1} \varphi_2 \land \mathbb{I}_{i@(\varphi_1\land \varphi_2)} \varphi_3)
\rightarrow \mathbb{I}_{i@\varphi_1} \varphi_3 
\\

\textbf{Ax}_{\mathsf{I3}}:\ \mathbb{I}_{i@(\mathbb{R}_i \varphi_1)} \varphi_2 \to \mathbb{I}_{i@ \varphi_1} \varphi_2 
&
\textbf{Ax}_{\mathsf{I6}}:\ (\mathbb{I}_{i@\varphi_1} \varphi_2 \land \mathbb{I}_{i@\varphi_1} (\varphi_2 \rightarrow \varphi_3))
\rightarrow \mathbb{I}_{i@\varphi_1} \varphi_3
\\[0.9em]
\textbf{Ax}_{\mathsf{B1}}:\ \mathbb{B}_{i@\varphi_1} \varphi_2 \leftrightarrow (\mathbb{R}_i \varphi_1 \land \mathbb{I}_{i@\varphi_1} \varphi_2)
&
\textbf{Ax}_{\mathsf{B2}}:\ \mathbb{B}_{i@\varphi_1} \varphi_2 \rightarrow \mathbb{R}_i (\varphi_1 \land \varphi_2)
\\[0.9em]

\textbf{Ax}_{\mathsf{S1}}:\ (\varphi_1 \land \mathbb{B}_{i@\varphi_1} \varphi_2) \rightarrow \mathbb{S}_i \varphi_2
&
\textbf{Ax}_{\mathsf{S3}}:\  \mathbb{S}_i \varphi \leftrightarrow \mathbb{S}_i \mathbb{S}_i \varphi
\\

\textbf{Ax}_{\mathsf{S2}}:\  \mathbb{S}_i \varphi
\to \varphi
&
\textbf{Ax}_{\mathsf{S4}}:\ (\mathbb{S}_i \varphi_1 \land \mathbb{S}_i \varphi_2)
\leftrightarrow \mathbb{S}_i (\varphi_1 \land \varphi_2)
\\[0.9em]

\textbf{Ax}_{\mathsf{G1}}:\ \mathbb{G}_{\varphi_1} \varphi_2 \rightarrow \mathbb{B}_{i@\varphi_1} \varphi_2
&
\textbf{Ax}_{\mathsf{G2}}:\ \mathbb{G}_{\varphi_1} \varphi_2 \rightarrow \mathbb{B}_{i@\varphi_1} \mathbb{G}_{\varphi_1} \varphi_2
\\[0.9em]

\textbf{Ax}_{\mathsf{C1}}:\ \mathbb{C} \varphi \rightarrow \varphi
&
\textbf{Ax}_{\mathsf{C2}}:\ \mathbb{C} \varphi \rightarrow \mathbb{S}_i \mathbb{C} \varphi
\end{array}
\]
We also take as primitive rules of inference:
$$\textbf{R}_{\mathsf{R}}: \ \frac{\varphi}{\mathbb{R}_i \varphi} \ \ \ \textbf{R}_{\mathsf{I}}: \ \frac{\varphi_2}{\mathbb{I}_{i@\varphi_1} \varphi_2}$$ 
$$\textbf{R}_{\mathsf{G}}: \ \frac{\varphi_3 \to \bigwedge_{i \in N} \mathbb{B}_{i@\varphi_1} \varphi_3 \text{ \ \ \ }  \varphi_3 \to \bigwedge_{i \in N} \mathbb{B}_{i@\varphi_1} \varphi_2}{\varphi_3 \to \mathbb{G}_{\varphi_1} \varphi_2} \ \ \ \textbf{R}_{\mathsf{C}}: \ \frac{\varphi_1 \to \bigwedge_{i \in N} \mathbb{S}_{i} \varphi_1 \text{ \ \ \ } \varphi_1 \to \varphi_2}{\varphi_1 \to \mathbb{C} \varphi_2}$$
In the next section, we show soundness of this proof system with respect to our semantics. We leave it as an open conjecture whether the proof system is also complete.

\section{Soundness}
We begin by showing soundness of each axiom schema.
\\
\\
$\textbf{Ax}_{\mathsf{R}}:\ \mathbb{R}_i \varphi \leftrightarrow \mathbb{R}_i \mathbb{R}_i \varphi$
\\
\underline{Proof}
\\
Let $\mathcal{F}=(\Omega,(\mathcal{E}_i,n_i)_{i \in N})$ be an arbitrary frame, $i$ be an agent in $N$, and $W$ be any subset of $\Omega$. It suffices to show $\textbf{R}_i(W)=\textbf{R}_i(\textbf{R}_i(W))$. Note that $\textbf{R}_i(W)=\{w \in \Omega : \exists E \in \mathcal{E}_i(w), \text{Yes}_i(W |E)\}$ is the union of all information states $E$ in $\mathcal{E}_i$ for which $\text{Yes}_i(W|E)$ holds. Hence, $\textbf{R}_i(W)$ is $1$-open in $\mathcal{T}_i$ and therefore also $(n_i+1)$-open. By Theorem 3.2, this means $\textbf{R}_i(W) \subseteq \textbf{R}_i(\textbf{R}_i(W))=\{w \in \Omega : \exists E \in \mathcal{E}_i(w), \text{Yes}_i(\textbf{R}_i(W)|E)\}$. We claim also that $\textbf{R}_i(\textbf{R}_i(W)) \subseteq \textbf{R}_i(W)$. So suppose $w \in \textbf{R}_i(\textbf{R}_i(W))$. Then $\exists E^* \in \mathcal{E}_i(w)$ so that $\text{Yes}_i(\textbf{R}_i(W)|E^*)$ holds, i.e. $\exists k \leq n_i$ so that $\textbf{R}_i(W) \cap E^*$ is $k$-closed but not $(k-1)$-open in $\mathcal{T}_i |E^*$.  Since $\textbf{R}_i(W)$ is $1$-open in $\mathcal{T}_i$, we know $\textbf{R}_i(W) \cap E^*$ is also $1$-open in $\mathcal{T}_i|E^*$. Therefore, it must be that $\textbf{R}_i(W) \cap E^*$ is $0$-closed in $\mathcal{T}_i|E^*$, i.e. $E^* \subseteq \textbf{R}_i(W)$. Thus, since $w \in E^*$, we conclude $w \in \textbf{R}_i(W)$. $\square$ 
\\
\\
$\textbf{Ax}_{\mathsf{I1}}:\ \mathbb{I}_{i@\varphi} \varphi$
\\
\underline{Proof}
\\
Let $\mathcal{F}=(\Omega,(\mathcal{E}_i,n_i)_{i \in N})$ be an arbitrary frame, $i$ be an agent in $N$, and $W$ be any subset of $\Omega$. It suffices to show $\forall w \in \Omega$ that $w \in \textbf{I}_{i@W}(W)$. Take arbitrary $w \in \Omega$. Note $\forall E \in \mathcal{E}_{i}(w), \text{Yes}_i(W|E) \to W \cap E \subseteq W$ as we always know $W \cap E \subseteq W$. By definition then, $w \in \textbf{I}_W(W)$. $\square$
\\
\\
$\textbf{Ax}_{\mathsf{I2}}:\ \mathbb{I}_{i@\varphi_1} \mathbb{I}_{i@\varphi_1} \varphi_2 \rightarrow \mathbb{I}_{i@\varphi_1} \varphi_2$
\\
\underline{Proof}
\\
Let $\mathcal{F}=(\Omega,(\mathcal{E}_i,n_i)_{i \in N})$ be an arbitrary frame, $i$ be an agent in $N$, and $W,P$ be any subsets of $\Omega$. It suffices to show $\textbf{I}_{i@W}(\textbf{I}_{i@W}(P)) \subseteq \textbf{I}_{i@W}(P)$. Take arbitrary $w \in \textbf{I}_{i@W}(\textbf{I}_{i@W}(P))$ and arbitrary $E^* \in \mathcal{E}_i(w)$. If $\text{Yes}_i(W|E^*)$ holds then $W \cap E^* \subseteq \textbf{I}_{i@W}(P)$. Then $\forall w' \in W \cap E^*$, since $w' \in \textbf{I}_{i@W}(P)$ and $E^* \in \mathcal{E}_i(w')$, it must be that $w' \in P$ and so $W \cap E^* \subseteq P$. Since $E^*$ was arbitrary, $\forall E \in \mathcal{E}_i(w), \text{Yes}_i(W|E) \to W \cap E \subseteq P$. By definition then, $w \in \textbf{I}_{i@W}(P)$. $\square$
\\
\\
$\textbf{Ax}_{\mathsf{I3}}:\ \mathbb{I}_{i@(\mathbb{R}_i \varphi_1)} \varphi_2 \to \mathbb{I}_{i@\varphi_1} \varphi_2$
\\
\underline{Proof}
\\
Let $\mathcal{F}=(\Omega,(\mathcal{E}_i,n_i)_{i \in N})$ be an arbitrary frame, $i$ be an agent in $N$, and $W,P$ be any subsets of $\Omega$. It suffices to show $\textbf{I}_{i@\textbf{R}_i(W)}(P) \subseteq \textbf{I}_{i@W}(P)$. Take arbitrary $w \in \textbf{I}_{i@\textbf{R}_i(W)}(P)$ and arbitrary $E^* \in \mathcal{E}_i(w)$. If $\text{Yes}_i(W|E^*)$ holds then we know $E^*\subseteq \textbf{R}_i(W)$, i.e. $\text{Yes}_i(\textbf{R}_i(W)|E^*)$ holds. Further, this means $\textbf{R}_i(W) \cap E^* \subseteq P$. As $E^* \subseteq \textbf{R}_i(W)$, it follows $E^* \subseteq P$ and so $W \cap E^* \subseteq P$. Since $E^*$ was arbitrary,  $\forall E \in \mathcal{E}_i(w), \text{Yes}_i(W|E) \to W \cap E \subseteq P$. By definition then, $w \in \textbf{I}_{i@W}(P)$.  $\square$
\\
\\
$\textbf{Ax}_{\mathsf{I4}}:\ \mathbb{I}_{i@\varphi_1} \varphi_2 \to \mathbb{I}_{i@\varphi_1} \mathbb{R}_i(\varphi_1 \land \varphi_2)$
\\
\underline{Proof}
\\
Let $\mathcal{F}=(\Omega,(\mathcal{E}_i,n_i)_{i \in N})$ be an arbitrary frame, $i$ be an agent in $N$, and $W,P$ be any subsets of $\Omega$. It suffices to show $\textbf{I}_{i@W}(P) \subseteq \textbf{I}_{i@W}\textbf{R}_i(W \cap P)$. Take arbitrary $w \in \textbf{I}_{i@W}(P) $ and arbitrary $E^* \in \mathcal{E}_i(w)$. If $\text{Yes}_i(W |E^*)$ holds then note $W \cap E^* \subseteq P$ and so $W \cap E^*=(W \cap P) \cap E^*$. Hence if $\text{Yes}_i(W|E^*)$ holds then $\text{Yes}_i(W \cap P|E^*)$ also holds. Take arbitrary $w' \in W\cap E^*$. Since $E^* \in \mathcal{E}_i(w')$ we know that $w' \in \textbf{R}_i(W \cap E^*)$.  Further, since $w'$ was arbitrary, this means $W \cap E^* \subseteq \textbf{R}_i(W \cap P)$. Finally, since $E^*$ was arbitrary, we have that  $\forall E \in \mathcal{E}_i(w), \text{Yes}_i(W|E) \to W \cap E \subseteq \textbf{R}_i(W \cap P)$. By definition then, $w \in \textbf{I}_{i@W}(\textbf{R}_i(W \cap P))$. $\square$
\\
\\
$\textbf{Ax}_{\mathsf{I5}}:\ (\mathbb{I}_{i@\varphi_1} \varphi_2 \land \mathbb{I}_{i@(\varphi_1\land \varphi_2)} \varphi_3)
\rightarrow \mathbb{I}_{i@\varphi_1} \varphi_3$
\\
\underline{Proof}
\\
Let $\mathcal{F}=(\Omega,(\mathcal{E}_i,n_i)_{i \in N})$ be an arbitrary frame, $i$ be an agent in $N$, and $W,P,Q$ be any subsets of $\Omega$. It suffices to show $\textbf{I}_{i@W}(P) \cap \textbf{I}_{i@(W\cap P)}(Q) \subseteq \textbf{I}_{i@W}(Q)$. Take arbitrary $w \in \textbf{I}_{i@W}(P) \cap \textbf{I}_{i@(W\cap P)}(Q)$ and arbitrary $E^* \in \mathcal{E}_i(w)$. If $\text{Yes}_i(W|E^*)$ holds then $W \cap E^* \subseteq P$ and so $W \cap E^*=(W\cap P) \cap E^*$. Hence, if $\text{Yes}_i(W|E^*)$ holds then $\text{Yes}_i(W\cap P |E^*)$ also holds and so $W \cap E^*=(W \cap P) \cap E^* \subseteq Q$. Since $E^*$ was arbitrary, this means  $\forall E \in \mathcal{E}_i(w), \text{Yes}_i(W|E) \to W \cap E \subseteq Q$. By definition then, $w \in \textbf{I}_{i@W}(Q)$. $\square$
\\
\\
$\textbf{Ax}_{\mathsf{I6}}:\ 
(\mathbb{I}_{i@\varphi_1} \varphi_2 \land \mathbb{I}_{i@\varphi_1} (\varphi_2 \rightarrow \varphi_3))
\rightarrow \mathbb{I}_{i@\varphi_1} \varphi_3$
\\
\underline{Proof}
\\
Let $\mathcal{F}=(\Omega,(\mathcal{E}_i,n_i)_{i \in N})$ be an arbitrary frame, $i$ be an agent in $N$, and $W,P,Q$ be any subsets of $\Omega$. It suffices to show $\textbf{I}_{i@W}(P) \cap \textbf{I}_{i@W}(P \to Q) \subseteq \textbf{I}_{i@W}(Q)$. Take arbitrary $w \in \textbf{I}_{i@W}(P) \cap \textbf{I}_{i@W}(P \to Q)$ and arbitrary $E^* \in \mathcal{E}_i(w)$. If $\text{Yes}_i(W|E^*)$ holds then $W \cap E^* \subseteq P$ and $W \cap E^* \subseteq (\Omega \backslash P) \cup Q$. Hence, $W \cap E^* \subseteq P \cap ((\Omega \backslash P) \cup Q)=Q$. Since $E^*$ was arbitrary, this means  $\forall E \in \mathcal{E}_i(w), \text{Yes}_i(W|E) \to W \cap E \subseteq Q$. By definition then, $w \in \textbf{I}_{i@W}(Q)$. $\square$
\\
\\
$\textbf{Ax}_{\mathsf{B1}}:\ \mathbb{B}_{i@\varphi_1} \varphi_2 \leftrightarrow (\mathbb{R}_i \varphi_1 \land \mathbb{I}_{i@\varphi_1} \varphi_2)$
\\
\underline{Proof}
\\
Let $\mathcal{F}=(\Omega,(\mathcal{E}_i,n_i)_{i \in N})$ be an arbitrary frame, $i$ be an agent in $N$, and $W,P$ be any subsets of $\Omega$. It suffices to show $\textbf{B}_{i@W}(P)=\textbf{R}_i(W) \cap \textbf{I}_{i@W}(P)$, which holds by definition. $\square$
\\
\\
$\textbf{Ax}_{\mathsf{B2}}:\ \mathbb{B}_{i@\varphi_1} \varphi_2 \rightarrow \mathbb{R}_i (\varphi_1 \land \varphi_2)$
\\
\underline{Proof}
\\
Let $\mathcal{F}=(\Omega,(\mathcal{E}_i,n_i)_{i \in N})$ be an arbitrary frame, $i$ be an agent in $N$, and $W,P$ be any subsets of $\Omega$. It suffices to show $\textbf{B}_{i@W}(P) \subseteq \textbf{R}_i(W \cap P)$. Take arbitrary $w \in \textbf{B}_{i@W}(P)$. Then $w \in \textbf{R}_i(W)$ and so $\exists E^* \in \mathcal{E}_i(w)$ for which $\text{Yes}_i(W|E^*)$ holds. Since $w \in \textbf{I}_{i@W}(P)$, this means $W \cap E^* \subseteq P$ and thus $W \cap E^*=(W\cap P) \cap E^*$. Hence, $\text{Yes}_i(W\cap P |E^*)$ also holds and so $w \in \textbf{R}_i(W \cap P)$. $\square$
\\
\\
$\textbf{Ax}_{\mathsf{S1}}:\ (\varphi_1 \land \mathbb{B}_{i@\varphi_1} \varphi_2) \rightarrow \mathbb{S}_i \varphi_2$
\\
\underline{Proof}
\\
Let $\mathcal{F}=(\Omega,(\mathcal{E}_i,n_i)_{i \in N})$ be an arbitrary frame, $i$ be an agent in $N$, and $W,P$ be any subsets of $\Omega$. It suffices to show $W \cap \textbf{B}_{i@W}(P) \subseteq \textbf{S}_i(P)$, which holds by definition. $\square$
\\
\\
$\textbf{Ax}_{\mathsf{S2}}:\  \mathbb{S}_i \varphi \to \varphi$
\\
\underline{Proof}
\\
Let $\mathcal{F}=(\Omega,(\mathcal{E}_i,n_i)_{i \in N})$ be an arbitrary frame, $i$ be an agent in $N$, and $P$ be any subset of $\Omega$. It suffices to show $\textbf{S}_i(P) \subseteq P$. We proved exactly this in Theorem 4.2. $\square$
\\
\\
$\textbf{Ax}_{\mathsf{S3}}:\  \mathbb{S}_i \varphi \leftrightarrow \mathbb{S}_i \mathbb{S}_i \varphi$
\\
\underline{Proof}
\\
Let $\mathcal{F}=(\Omega,(\mathcal{E}_i,n_i)_{i \in N})$ be an arbitrary frame, $i$ be an agent in $N$, and $P$ be any subset of $\Omega$. It suffices to show $\textbf{S}_i(P)=\textbf{S}_i(\textbf{S}_i(P))$. We proved exactly this in Theorem 4.3. $\square$
\newpage
\noindent $\textbf{Ax}_{\mathsf{S4}}:\ (\mathbb{S}_i \varphi_1 \land \mathbb{S}_i \varphi_2)
\leftrightarrow \mathbb{S}_i (\varphi_1 \land \varphi_2)$
\\
\underline{Proof}
\\
Let $\mathcal{F}=(\Omega,(\mathcal{E}_i,n_i)_{i \in N})$ be an arbitrary frame, $i$ be an agent in $N$, and $P_1,P_2$ be any subsets of $\Omega$. It suffices to show $\textbf{S}_i(P_1) \cap \textbf{S}_i(P_2) =\textbf{S}_i(P_1 \cap P_2)$. We proved exactly this in Theorem 4.4. $\square$
\\
\\
\noindent $\textbf{Ax}_{\mathsf{G1}}:\ \mathbb{G}_{\varphi_1} \varphi_2 \rightarrow \mathbb{B}_{i@\varphi_1} \varphi_2$
\\
\underline{Proof}
\\
Let $\mathcal{F}=(\Omega,(\mathcal{E}_i,n_i)_{i \in N})$ be an arbitrary frame, $i$ be an agent in $N$, and $W,P$ be any subsets of $\Omega$. It suffices to show $\textbf{G}_{W}(P) \subseteq \textbf{B}_{i@W}(P)$, which holds by definition. $\square$
\\
\\
$\textbf{Ax}_{\mathsf{G2}}:\ \mathbb{G}_{\varphi_1} \varphi_2 \rightarrow \mathbb{B}_{i@\varphi_1} \mathbb{G}_{\varphi_1} \varphi_2$
\\
\underline{Proof}
\\
Let $\mathcal{F}=(\Omega,(\mathcal{E}_i,n_i)_{i \in N})$ be an arbitrary frame, $i$ be an agent in $N$, and $W,P$ be any subsets of $\Omega$. It suffices to show  $\textbf{G}_W(P) \subseteq \textbf{B}_{i@W}(\textbf{G}_W(P))$. Take arbitrary $w \in \textbf{G}_W(P)=\bigcap_{k \in \mathbb{N^+}} \textbf{E}_{W}^k(P)$. Since we know $w \in \textbf{E}_{W}^1(P) = \bigcap_{j \in N} \textbf{B}_{j@W}(P)$ it follows $w \in \textbf{B}_{i@W}(P) \subseteq \textbf{R}_{i}(W)$. Additionally, since $\forall k \in \mathbb{N}^+$ we know $w \in \textbf{E}^{k+1}_W(P)=\bigcap_{j \in N} \textbf{B}_{j@W}(\textbf{E}^{k}_W(P))$ it follows $w \in \textbf{B}_{i@W}(\textbf{E}^{k}_W(P)) \subseteq \textbf{I}_{i@W}(\textbf{E}^{k}_W(P))$. Take arbitrary $E^* \in \mathcal{E}_i(w)$. If $\text{Yes}_i(W|E^*)$ holds then $\forall k \in \mathbb{N}^+$ we know $W \cap E^* \subseteq \textbf{E}^{k}_W(P)$, i.e. we have $W \cap E^* \subseteq \textbf{G}_W(P)$. Hence, as $E^*$ was arbitrary, $w \in \textbf{I}_{i@W}(\textbf{G}_W(P))$ and so  $w \in \textbf{B}_{i@W}(\textbf{G}_W(P))$. $\square$
\\
\\
$\textbf{Ax}_{\mathsf{C1}}:\ \mathbb{C} \varphi \rightarrow \varphi$
\\
\underline{Proof}
\\
Let $\mathcal{F}=(\Omega,(\mathcal{E}_i,n_i)_{i \in N})$ be an arbitrary frame and $P$ be any subset of $\Omega$. It suffices to show $\textbf{C}(P) \subseteq P$. Recall $\textbf{C}(P)$ is a fixed point of the map $X \mapsto P \cap \bigcap_{i \in N} \textbf{S}_i(X)$. Thus, $\textbf{C}(P)= P \cap \bigcap_{i \in N} \textbf{S}_i(\textbf{C}(P)) \subseteq  P$ . $\square$
\\
\\
$\textbf{Ax}_{\mathsf{C2}}:\ \mathbb{C} \varphi \rightarrow \mathbb{S}_i \mathbb{C} \varphi$
\\
\underline{Proof}
\\
Let $\mathcal{F}=(\Omega,(\mathcal{E}_i,n_i)_{i \in N})$ be an arbitrary frame, $i$ be an agent in $N$, and $P$ be any subset of $\Omega$. It suffices to show $\textbf{C}(P) \subseteq \textbf{S}_i(\textbf{C}(P))$. Recall $\textbf{C}(P)$ is a fixed point of the map $X \mapsto P \cap \bigcap_{j \in N} \textbf{S}_j(X)$. Thus, $\textbf{C}(P)= P \cap \bigcap_{j \in N} \textbf{S}_j(\textbf{C}(P)) \subseteq \textbf{S}_i(\textbf{C}(P))$ as desired. $\square$
\\
\\
We now show soundness of each inference rule.
\\
\\
$\textbf{R}_{\mathsf{R}}: \ \frac{\varphi}{\mathbb{R}_i \varphi}$
\\
\underline{Proof}
\\
Let $\mathcal{F}=(\Omega,(\mathcal{E}_i,n_i)_{i \in N})$ be an arbitrary frame and $i$ be an agent in $N$. It suffices to show $\forall w \in \Omega$ that $w \in \textbf{R}_i(\Omega)$. Take arbitrary $w \in \Omega$. Then there exists some $E^* \in \mathcal{E}_i(w)$. Further, since $\Omega \cap E^*=E^*$ is $0$-closed in $\mathcal{T}_i |E^*$, we know $\text{Yes}_i(W|E^*)$ holds. Therefore $w \in \textbf{R}_i(\Omega)$. $\square$
\\
\\
$\textbf{R}_{\mathsf{I}}: \ \frac{\varphi_2}{\mathbb{I}_{i@\varphi_1} \varphi_2}$
\\
\underline{Proof}
\\
Let $\mathcal{F}=(\Omega,(\mathcal{E}_i,n_i)_{i \in N})$ be an arbitrary frame and $i$ be an agent in $N$ and $W$ be any subset of $\Omega$. It suffices to show $\forall w \in \Omega$ that $w \in \textbf{I}_{i@W}(\Omega)$. Note $\forall E \in \mathcal{E}_{i}(w), \text{Yes}_i(W|E) \to W \cap E \subseteq \Omega$ as we always know $W \cap E \subseteq \Omega$. By definition then, $w \in \textbf{I}_W(\Omega)$. $\square$
\\
\\
To simplify our argument as to why $\textbf{R}_{\mathsf{G}}$ is sound, we next prove two key lemmas... 
\newpage
\noindent \textbf{Lemma 6.0:} Let $\mathcal{F}=(\Omega,(\mathcal{E}_i,n_i)_{i \in N})$ be an arbitrary frame and $W,P$ be any subsets of $\Omega$. Then the map $X \mapsto \bigcap_{i \in N} \left(\textbf{B}_{i@W} (P)\cap \textbf{B}_{i@W} (X)\right)$  is monotone.
\\
\underline{Proof}
\\
Suppose $X_1 \subseteq X_2$. Take arbitrary $w \in \bigcap_{i \in N} \left(\textbf{B}_{i@W} (P)\cap \textbf{B}_{i@W} (X_1)\right)$. Since $\forall i \in N$ we know $w \in \textbf{B}_{i@W}(P)$ it follows $\forall i \in N$ that $w \in \textbf{R}_i(W)$. Take arbitrary $E^* \in \mathcal{E}_i(w)$. If $\text{Yes}_i(W|E^*)$ holds then $W \cap E^* \subseteq X_1$ and hence  $W \cap E^* \subseteq X_2$. Hence, as $E^*$ was arbitrary, we have $\forall i \in N$ that $w \in \textbf{I}_{i@W}(X_2)$. Thus, $\forall i \in N$ it follows $w \in \textbf{B}_{i@W}(X_2)$. In particular, $w \in \bigcap_{i \in N} \left(\textbf{B}_{i@W} (P)\cap \textbf{B}_{i@W} (X_2)\right)$. Therefore, the map $X \mapsto \bigcap_{i \in N} \left(\textbf{B}_{i@W} (P)\cap \textbf{B}_{i@W} (X)\right)$ is monotone. $\square$
\\
\\
\textbf{Lemma 6.1:} Let $\mathcal{F}=(\Omega,(\mathcal{E}_i,n_i)_{i \in N})$ be an arbitrary frame and $W,P$ be any subsets of $\Omega$. Then the map $X \mapsto \bigcap_{i \in N} \left(\textbf{B}_{i@W} (P)\cap \textbf{B}_{i@W} (X)\right)$  commutes with intersection.
\\
\underline{Proof}
\\
Let $(X_j)_{j \in J}$ be an arbitrary collection of subsets of $\Omega$. Note first that $\bigcap_{j \in J} \bigcap_{i \in N} \left(\textbf{B}_{i@W} (P) \cap \textbf{B}_{i@W} (X_j)\right)=\bigcap_{i \in N} (\textbf{B}_{i@W}(P) \cap \bigcap_{j \in J} \textbf{B}_{i@W}(X_j))=\bigcap_{i \in N} (\textbf{B}_{i@W}(P) \cap \bigcap_{j \in J} \textbf{I}_{i@W}(X_j))$. Now observe $w \in \cap_{j \in J} \textbf{I}_{i@W}(X_j)$ iff $\forall j \in J$ we have $\forall E \in \mathcal{E}_{i}(w), \text{Yes}_i(W|E) \to W \cap E \subseteq X_j$. Therefore, $w  \in \bigcap_{j \in J} \textbf{I}_{i@W}(X_j)$ just in case it holds that $\forall E \in \mathcal{E}_{i}(w), \text{Yes}_i(W|E) \to W \cap E \subseteq \cap_{j \in J} X_j$, i.e. iff $w \in \textbf{I}_{i@W}(\cap_{j \in J} X_j)$. Hence, we have $\bigcap_{j \in J} \bigcap_{i \in N} \left(\textbf{B}_{i@W} (P) \cap \textbf{B}_{i@W} (X_j)\right)=\bigcap_{i \in N} (\textbf{B}_{i@W}(P) \cap \textbf{I}_{i@W}\left(\cap_{j \in J} X_j\right))=\bigcap_{i \in N} (\textbf{B}_{i@W}(P) \cap \textbf{B}_{i@W}\left(\cap_{j \in J} X_j\right))$. Thus, the map $X \mapsto \bigcap_{i \in N} \left(\textbf{B}_{i@W} (P)\cap \textbf{B}_{i@W} (X)\right)$ commutes with intersection. $\square$
\\
\\
\noindent $\textbf{R}_{\mathsf{G}}: \ \frac{\varphi_3 \to \bigwedge_{i \in N} \mathbb{B}_{i@\varphi_1} \varphi_3 \text{ \ \ \ }  \varphi_3 \to \bigwedge_{i \in N} \mathbb{B}_{i@\varphi_1} \varphi_2}{\varphi_3 \to \mathbb{G}_{\varphi_1} \varphi_2}$
\\
\underline{Proof}
\\
Let $\mathcal{F}=(\Omega,(\mathcal{E}_i,n_i)_{i \in N})$ be an arbitrary frame and $W,P,Q$ be any subsets of $\Omega$. It suffices to show  if $Q \subseteq \bigcap_{i \in  N} \textbf{B}_{i@W}(Q)$ and $Q \subseteq  \bigcap_{i \in  N} \textbf{B}_{i@W}(P)$ then $Q \subseteq \textbf{G}_W(P)$. By Lemma 6.0, Tarski's implies the map $X \mapsto \bigcap_{i \in N} \left(\textbf{B}_{i@W} (P)\cap \textbf{B}_{i@W} (X)\right)$ has a greatest (post)fixed point; by Lemma 6.1, if we let $\textbf{e}^k_W(P)$ be the result of applying the map $X \mapsto \bigcap_{i \in N} \left(\textbf{B}_{i@W} (P)\cap \textbf{B}_{i@W} (X)\right)$ iteratively $k$ times on $\Omega$, Kleene's theorem implies this greatest (post)fixed point is given by $\bigcap_{k \in \mathbb{N}^+} \textbf{e}^k_W(P)$. Observe, by monotonicity, that $\Omega \supseteq \textbf{e}^1_W(P) \supseteq \textbf{e}^2_W(P)...$ forms a descending sequence of subsets of $\Omega$. We will now show $\forall k \in \mathbb{N}^+$ that $\textbf{e}^k_W(P)=\textbf{E}^k_W(P)$ and so $\textbf{G}_W(P)=\bigcap_{k \in \mathbb{N}^+} \textbf{e}^k_W(P)$. We proceed by induction on $k$. When $k=1$, we know $\textbf{e}^1_W(P)=\bigcap_{i \in N} \left(\textbf{B}_{i@W} (P)\cap \textbf{B}_{i@W} (P)\right)=\textbf{E}^1_W(P)$ establishing the base case. For the inductive step, assume $\forall k \leq t$ we have $\textbf{e}^k_W(P)=\textbf{E}^k_W(P)$. Note $\textbf{e}^{t+1}_W(P)=\bigcap_{i \in N} \left(\textbf{B}_{i@W} (P)\cap \textbf{B}_{i@W} (\textbf{e}^t_W(P))\right)=\bigcap_{i \in N} \left(\textbf{B}_{i@W} (P)\cap \textbf{B}_{i@W} (\textbf{E}^t_W(P))\right)=\textbf{E}^1_W(P) \cap \textbf{E}^{t+1}_W(P)$. Take arbitrary $w \in \textbf{E}^{t+1}_W(P)=\bigcap_{i \in N} \textbf{B}_{i@W}(\textbf{E}^{t}_W(P))$. Then $\forall i \in N$ we have $w \in \textbf{R}_i(W)$ and $w\in \textbf{I}_{i@W}(\textbf{E}^{t}_W(P))$. Take arbitrary $E^* \in \mathcal{E}_i(w)$. If $\text{Yes}_i(W|E^*)$ holds then $W \cap E^* \subseteq \textbf{E}^{t}_W(P)$. By hypothesis, we know $\textbf{E}^{t}_W(P) \subseteq ... \subseteq \textbf{E}^{1}_W(P)$. Hence, if $\text{Yes}_i(W|E^*)$ holds then $W \cap E^* \subseteq \textbf{E}^{1}_W(P)$. Namely $\forall w' \in W \cap E^*$, since $w' \in \textbf{E}^{1}_W(P) \subseteq \textbf{I}_{i@W}(P)$ and $E^* \in \mathcal{E}_i(w')$, it must be that $w' \in P$ and so $W \cap E^* \subseteq P$. Hence, $w \in \textbf{I}_{i@W}(P)$ and therefore $w \in \textbf{E}^1_W(P)= \bigcap_{i \in N}\textbf{B}_{i@W}(P)$. In particular, as $\textbf{E}^{t+1}_W(P) \subseteq \textbf{E}^1_W(P)$, this means $\textbf{e}^{t+1}_W(P)=\textbf{E}^{t+1}_W(P)$ thereby completing the inductive step. Since $Q \subseteq \bigcap_{i \in  N} \left(\textbf{B}_{i@W}(P) \cap \textbf{B}_{i@W}(Q)\right)$, it is a postfixed point of $X \mapsto \bigcap_{i \in N} \left(\textbf{B}_{i@W} (P)\cap \textbf{B}_{i@W} (X)\right)$. As the greatest (post)fixed point of $X \mapsto \bigcap_{i \in N} \left(\textbf{B}_{i@W} (P)\cap \textbf{B}_{i@W} (X)\right)$  is $\textbf{G}_W(P)$, it follows $Q \subseteq \textbf{G}_W(P)$. $\square$
\\
\\
$\textbf{R}_{\mathsf{C}}: \ \frac{\varphi_1 \to \bigwedge_{i \in N} \mathbb{S}_{i} \varphi_1 \text{ \ \ \ } \varphi_1 \to \varphi_2}{\varphi_1 \to \mathbb{C} \varphi_2}$
\\
\underline{Proof} 
\\
Let $\mathcal{F}=(\Omega,(\mathcal{E}_i,n_i)_{i \in N})$ be an arbitrary frame and $W,P$ be any subsets of $\Omega$. It suffices to show  if $W \subseteq \bigcap_{i \in  N} \textbf{S}_i(W)$ and $W \subseteq P$ then $W \subseteq \textbf{C}(P)$. Since $W \subseteq P \cap \bigcap_{i \in  N} \textbf{S}_i(W)$, by Theorem 4.2, we know $W=P \cap \bigcap_{i \in  N} \textbf{S}_i(W)$, i.e. it is a fixed point of $X \mapsto P \cap \bigcap_{i \in N} \textbf{S}_i(X)$. By definition then, $W \subseteq \textbf{C}(P)$. $\square$

\section{Inductive Coordinated Attack}
Consider the following inductive variant of the coordinated attack problem:\footnote[11]{Halpern \& Moses 1990 were the first to analyze coordinated attack using a logic for common knowledge. To our awareness, this variant (despite its clear applications for designing distributed consensus protocols) has not received any such treatment.}
\begin{quote}
Each agent $i \in N$ is deciding whether to attest $P$ is true or defer judgement. Let $n_i$ be the maximum number of mind changes agent $i$ is willing to tolerate after first attesting $P$ is true.
\begin{itemize}
\item If $P$ is true, each agent $i$ ideally wants everyone to converge on attesting $P$ is true.
\item However, if $P$ is false or some agent $i$ does not converge on attesting $P$ is true, then anyone who converges on attesting $P$ is true is perpetually shamed.
\end{itemize}
What non-trivial attestation protocols are available to the agents, if any, so that they can guarantee none of them will be perpetually shamed?
\end{quote}
More formally, suppose we have a frame $\mathcal{F}=(\Omega,(\mathcal{E}_i,n_i)_{i \in N})$. An \emph{attestation strategy} for $i$ is a map $s_i: \mathcal{E}_i \to \{\text{Yes}, \text{?}\}$ such that the decision method $m_i:  \mathcal{E}_i \to \{\text{Yes}, \text{No}\}$ which $\forall E \in \mathcal{E}_i$ sets $m_i(E)=\text{Yes}$ iff $s_i(E)=\text{Yes}$ has at most $n_i$ switches after saying `Yes.' Additionally, as with decision methods, define $\sigma_{s_i}: \Omega \to \{\text{Yes}, \text{?}\}$ to be the function which for each $w \in \Omega$ returns $\sigma_{s_i}(w)$ as the output $s_i$ converges to in world $w$. Finally, an \emph{attestation protocol} for $N$ is a tuple $(s_i)_{i \in N}$ of attestation strategies for each $i \in N$. 
\\
\\
We say an attestation protocol $(s_i)_{i \in N}$  \emph{solves coordinated attack for} $P$ iff it satisfies:
\begin{enumerate}
\item \textbf{Validity-} $\forall i \in N$, $\sigma^{-1}_{s_i}(\text{Yes}) \subseteq P$.
\item \textbf{Agreement-} $\forall i,j \in N$, $\sigma^{-1}_{s_i}(\text{Yes})=\sigma^{-1}_{s_j}(\text{Yes})$.
\item \textbf{Nontriviality-} $\bigcap_{i \in N} \sigma^{-1}_{s_i}(\text{Yes}) \neq \varnothing$.
\end{enumerate}
Validity ensures that agents will never falsely converge on attesting $P$ is true. Agreement ensures that agents will all converge on the same output in every world. Validity and agreement together ensure that no one is perpetually shamed. A trivial way to satisfy validity and agreement is to take the attestation protocol where each agent always defers judgement on $P$. To rule out such pathological protocols, the nontriviality condition further ensures there is at least one world where everyone converges on attesting $P$ is true. In particular, given an attestation protocol $(s_i)_{i\in N}$ which solves consensus for $P$, we will say that $\bigcap_{i \in N} \sigma^{-1}_{s_i}(\text{Yes})$ is the \emph{success set} of $(s_i)_{i\in N}$.
\\
\\
\textbf{Theorem 7.0:} There exists an attestation protocol $(s_i)_{i \in N}$ which solves coordinated attack for $P$ with success set $W$ if and only if $W$ is a non-empty $(n_i+1)$-open subset of $P$ in each $\mathcal{T}_i$.
\\
\underline{Proof}
\\
First we show the `if' direction. So suppose $W$ is a non-empty $(n_i+1)$-open subset of $P$ in each $\mathcal{T}_i$. By Theorem 3.0, $\forall i \in N$ there is a decision method $m_{i}$ has at most $n_i$ switches after saying `Yes' and $\sigma^{-1}_{m_{i}}(\text{Yes})=W$. Consider the attestation protocol $(s_i)_{i \in N}$ which $\forall i \in N$ sets $s_i(E)=\text{Yes}$ iff $m_{i}(E)=\text{Yes}$. Clearly, $\forall i \in N$ we have $\sigma^{-1}_{s_i}(\text{Yes})=\sigma^{-1}_{m_{i}}(\text{Yes})=W \subseteq P$. Hence, because $W$ is non-empty, it immediately follows that $(s_i)_{i \in N}$ solves coordinated attack for $P$ with success set $W$.
\\
Now we show the `only if' direction. Suppose there exists an attestation protocol $(s_i)_{i \in N}$ which solves coordinated attack for $P$ with success set $W$. Agreement guarantees $\forall i \in N$ that $\sigma^{-1}_{s_i}(\text{Yes})=W$. Non-triviality and validity further guarantee $W$ is a non-empty subset of $P$. Additionally, $\forall i \in N$ the decision method $m_i:  \mathcal{E}_i \to \{\text{Yes}, \text{No}\}$ which sets $m_i(E)=\text{Yes}$ iff $s_i(E)=\text{Yes}$ has at most $n_i$ switches after saying `Yes.' Since $\sigma_{m_i}^{-1}(\text{Yes})=\sigma^{-1}_{s_i}(\text{Yes})=W$, this means $W$ is $(n_i+1)$-open in each $\mathcal{T}_i$. $\square$
\newpage
\noindent Theorem 7.0 implies that the possible success sets of an attestation protocol solving coordinated attack for $P$ are precisely those non-empty subsets of $P$ which are $(n_i+1)$-open in each $\mathcal{T}_i$. Additionally, given any such subset $W$ of $P$, Theorem 3.0 allows us to explicitly construct at least one attestation protocol solving coordinated attack for $P$ with success set $W$. In effect then, if one regards two attestation protocols solving coordinated attack for $P$ as equally successful whenever they have the same success sets, the problem of solving coordinated attack for $P$ can be reduced to the task of simply finding those non-empty subsets of $P$ which are $(n_i+1)$-open in each $\mathcal{T}_i$.
\\
\\
\textbf{Theorem 7.1:} $W$ is a $(n_i+1)$-open subset of $P$ in each $\mathcal{T}_i$  if and only if $W$ is a fixed point of the map $X \mapsto X \cap \textbf{G}_X(P)$. Further, $\forall W \subseteq \Omega$, $W \cap \textbf{G}_W(P)$ is itself such a fixed point.
\\
\underline{Proof}
\\
Suppose $W$ is a $(n_i+1)$-open subset of $P$ in each $\mathcal{T}_i$. By our proof of $\textbf{Ax}_{\mathsf{I1}}$, we know $\forall i \in N$ that $\textbf{I}_{i@W}(W)=\Omega$. Moreover, by Theorem 3.2, we have $W \subseteq \textbf{R}_i(W)=\{w \in \Omega : \exists E \in \mathcal{E}_i(w), \text{Yes}_i(W|E)\}$. Hence, $W \subseteq \textbf{R}_i(W) \cap \textbf{I}_{i@W}(W)=\textbf{B}_{i@W}(W)$ and so $W \subseteq \bigcap_{i \in N} \textbf{B}_{i@W}(W)$. Next,  $\forall i \in N$ we have that $\textbf{I}_{i@W}(P)=\{w \in \Omega: \forall E \in \mathcal{E}_i(w), \text{Yes}_i(W|E) \to W \cap E \subseteq P\}=\Omega$ as $W \subseteq P$.   In particular, it follows $W \subseteq \textbf{R}_i(W) \cap \textbf{I}_{i@W}(P)=\textbf{B}_{i@W}(P)$ and so $W \subseteq \bigcap_{i \in N} \textbf{B}_{i@W}(P)$.  Therefore, using our proof of $\textbf{R}_{\mathsf{G}}$, this means $W \subseteq \textbf{G}_W(P)$, i.e. $W=W \cap \textbf{G}_{W}(P)$.
\\
Now suppose $W=W \cap \textbf{G}_{W}(P)$. By our proof of $\textbf{Ax}_{\mathsf{G1}}$, we know $\forall i \in N$ that $\textbf{G}_{W}(P) \subseteq \textbf{B}_{i@W}(P)$. Hence, $W \subseteq W \cap \textbf{B}_{i@W}(P)$. By our proof of $\textbf{Ax}_{\mathsf{S1}}$, we have $W \cap \textbf{B}_{i@W}(P) \subseteq \textbf{S}_i(P)$. Further, by Theorem 4.2, $ \textbf{S}_i(P)\subseteq P$. Therefore, $W \subseteq P$. Also by definition, we know that $\textbf{B}_{i@W}(P) \subseteq \textbf{R}_i(W)$. Hence, $W \subseteq \textbf{R}_i(W)=\{w \in \Omega : \exists E \in \mathcal{E}_i(w), \text{Yes}_i(W|E)\}$ and so $W$ is $(n_i+1)$-open in $\mathcal{T}_i$ by Theorem 3.2. Thus, $W$ is a $(n_i+1)$-open subset of $P$ in each $\mathcal{T}_i$.
\\
Finally consider arbitrary $W \subseteq \Omega$. We claim $W \cap \textbf{G}_W(P)=(W \cap \textbf{G}_W(P)) \cap \textbf{G}_{W \cap \textbf{G}_W(P)}(P)$. By our proof of $\textbf{Ax}_{\mathsf{G2}}$, we know $\forall i \in N$ that $\textbf{G}_{W}(P) \subseteq \textbf{B}_{i@W}(\textbf{G}_{W}(P))$. Also, we know $\textbf{B}_{i@W}(\textbf{G}_{W}(P)) \subseteq \textbf{R}_i(W \cap \textbf{G}_{W}(P))$ by our proof of $\textbf{Ax}_{\mathsf{B2}}$. Thus, $W \cap \textbf{G}_W(P) \subseteq \textbf{R}_i(W \cap \textbf{G}_{W}(P))$. Next, by our proof of $\textbf{Ax}_{\mathsf{I1}}$, we know $\forall i \in N$ that $\textbf{I}_{i@(W\cap \textbf{G}_{W}(P))}(W \cap \textbf{G}_{W}(P))=\Omega$. Since $\textbf{R}_i(W \cap \textbf{G}_{W}(P)) \cap \textbf{I}_{i@(W\cap \textbf{G}_{W}(P))}(W \cap \textbf{G}_{W}(P))=\textbf{B}_{i@(W \cap \textbf{G}_{W}(P))}(W \cap \textbf{G}_{W}(P))$, we have $W \cap \textbf{G}_W(P) \subseteq \textbf{B}_{i@(W \cap \textbf{G}_{W}(P))}(W \cap \textbf{G}_{W}(P))$. In particular, this means $W \cap \textbf{G}_W(P) \subseteq \bigcap_{i \in N} \textbf{B}_{i@(W \cap \textbf{G}_{W}(P))}(W \cap \textbf{G}_{W}(P))$. By our proof of $\textbf{Ax}_{\mathsf{B1}}$, we know $\textbf{G}_{W}(P) \subseteq \textbf{B}_{i@W}(P)$ and so $W \cap \textbf{G}_{W}(P) \subseteq W \cap \textbf{B}_{i@W}(P)$. By our proof of $\textbf{Ax}_{\mathsf{S1}}$, $W \cap \textbf{B}_{i@W}(P) \subseteq \textbf{S}_i(P)$. Further, by Theorem 4.2, we have $\textbf{S}_i(P) \subseteq P$. Thus, $W \cap \textbf{G}_W(P) \subseteq P$. Using this fact then, $\forall i \in N$ it follows that we have $\textbf{I}_{i@(W\cap \textbf{G}_{W}(P))}(P)=\{w \in \Omega: \forall E \in \mathcal{E}_i(w), \text{Yes}_i(W\cap \textbf{G}_{W}(P)|E) \to (W\cap \textbf{G}_{W}(P)) \cap E \subseteq P\}=\Omega$. Since $\textbf{R}_i(W \cap \textbf{G}_{W}(P)) \cap \textbf{I}_{i@(W\cap \textbf{G}_{W}(P))}(P)=\textbf{B}_{i@(W \cap \textbf{G}_{W}(P))}(P)$, we know $W \cap \textbf{G}_{W}(P) \subseteq \textbf{B}_{i@(W \cap \textbf{G}_{W}(P))}(P)$. In particular, this means $W \cap \textbf{G}_{W}(P) \subseteq \bigcap_{i \in N} \textbf{B}_{i@(W \cap \textbf{G}_{W}(P))}(P)$. Therefore, appealing to our proof of $\textbf{R}_{\mathsf{G}}$, we have $W \cap \textbf{G}_W(P) \subseteq \textbf{G}_{W \cap \textbf{G}_W(P)}(P)$ and so $W \cap \textbf{G}_W(P) = (W \cap \textbf{G}_W(P)) \cap \textbf{G}_{W \cap \textbf{G}_W(P)}(P)$. $\square$
\\
\\
Hence, we can use the map $X \mapsto X \cap \textbf{G}_X(P)$ to pick out any subset of $P$ which is $(n_i+1)$-open in each $\mathcal{T}_i$ and thereby construct (modulo success sets) all those attestation protocols which solve coordinated attack for $P$. Additionally, if one defines the operator $\textbf{L}: 2^{\Omega} \to 2^{\Omega}$ by $\forall P \subseteq \Omega$ setting:
$$\textbf{L}(P)=\{w \in \Omega : \exists W \subseteq \Omega, w \in W \cap \textbf{G}_W(P)\}$$
then $\textbf{L}(P)$ corresponds to those worlds which belong to the success set of some attestation protocol that solves coordinated attack for $P$. Note these are also exactly those worlds where some state of affairs holds such that it generates common inductive knowledge of $P$! Thus, $\textbf{L}(P)$ further corresponds to those worlds where $P$ can become Lewisian common inductive knowledge. As we argued in the pre-theory section, Lewis' definition of common inductive knowledge entails ours. In what follows, we provide a  simplified proof of this fact before using it to analyze inductive coordinated attack. 
\newpage
\noindent \textbf{Lemma 7.0:} $\textbf{L}(P) \subseteq \textbf{C}(P)$.
\\
\underline{Proof}
\\
Suppose $w \in \textbf{L}(P)$. Then $\exists W \subseteq \Omega$ for which $w \in W \cap \textbf{G}_W(P)$. By Theorem 7.1, $\forall i \in N$ we have $W \cap \textbf{G}_W(P)$ is $(n_i+1)$-open in each $\mathcal{T}_i$. Hence, by Lemma 4.2, $W \cap \textbf{G}_W(P) \subseteq \textbf{S}_i(W \cap \textbf{G}_W(P))$. In particular, $W \cap \textbf{G}_W(P) \subseteq \bigcap_{i \in N} \textbf{S}_i(W \cap \textbf{G}_W(P))$. While proving soundness of $\textbf{Ax}_{\mathsf{G1}}$, we showed that $\forall i \in N$ we have $\textbf{G}_W(P) \subseteq \textbf{B}_{i@W}(P)$. Hence, $W \cap \textbf{G}_W(P) \subseteq W \cap \textbf{B}_{i@W}(P)$. By our proof of $\textbf{Ax}_{\mathsf{S1}}$, we know   $W \cap \textbf{B}_{i@W}(P) \subseteq \textbf{S}_i(P)$. Further, by Theorem 4.2, $ \textbf{S}_i(P)\subseteq P$. Thus, $W \cap \textbf{G}_W(P) \subseteq P$. Appealing to our proof of $\textbf{R}_{\mathsf{C}}$, this means $W \cap \textbf{G}_W(P) \subseteq \textbf{C}(P)$ and therefore $w \in \textbf{C}(P)$ as desired. $\square$
\\
\\
By Lemma 7.0, since the success set of any attestation protocol which solves coordinated attack for $P$ is a subset of $\textbf{L}(P)$, we have that $\textbf{C}(P)$ is also an upper bound for any such success set. This upper bound may be unattainable as illustrated by our example on page 21 where $\textbf{L}(P) \subset \textbf{C}(P)$. Nevertheless, if $\Omega$ is finite and each $n_i$ is sufficiently high, then whenever $\textbf{C}(P)$ is non-empty we can show that there is an attestation protocol solving coordinated attack for $P$ whose success set attains this upper bound. Hence, asymptotically speaking, an attestation protocol solving coordinated attack for $P$ is `welfare-maximizing' iff its success set is $\textbf{C}(P)$ as the worlds outside $\textbf{C}(P)$ do not belong to the success set of any such attestation protocol. We now prove an important lemma necessary to establish this result...
\\
\\
\textbf{Lemma 7.1:} Suppose $\Omega$ is finite. For each $i$ there exists an $n$ so that $\textbf{C}(P)$ is $(n+1)$-open in $\mathcal{T}_i$. In particular, we can define $n^*_{i,P}$ to be the least such $n$ for every agent $i \in N$. 
\\
\underline{Proof}
\\
Since $\Omega$ is finite $\mathcal{T}_i$ only has finitely many 2-open sets. By Theorem 4.5 then, $\textbf{C}(P)$ is a finite union of $2$-open sets in $\mathcal{T}_i$. Let $W_0=O_0 \backslash O'_0,...,W_t=O_t \backslash O'_t$ be a finite sequence of $2$-open sets where each $O_k,O'_k \in \mathcal{T}_i$ and $O_k \supseteq O'_k$ so that $\textbf{C}(P)=\bigcup_{k \leq t} W_k$. Consider the decision method $m_i: \mathcal{E}_i \to \{\text{Yes}, \text{No}\}$ which $\forall E \in \mathcal{E}_i$ sets:
$$m_i(E)=\begin{cases} \text{Yes} & \exists k \leq t \text{ such that } E \subseteq O_k \text{ and } W_k \cap E \neq \varnothing \\ \text{No} & \text{else} \end{cases}$$	
We will first show $m_i$ has at most $2t+1$ switches after saying `Yes.' Towards that end, suppose for sake of contradiction there exists a $(2t+2)$-switching sequence $E_0 \supseteq E'_0 \supseteq ... E_t \supseteq E'_t \supseteq E_{t+1}$ for $m_i$ starting from `Yes.'  Then $\forall s \leq t+1$, because $m_i(E_s)=\text{Yes}$, there $\exists k_s \leq t$ such that $E_s \subseteq O_{k_s}$ and $W_{k_s} \cap E_s \neq \varnothing$. Further, since $m_i(E'_s)=\text{No}$ and $E'_s \subseteq E_s \subseteq O_{k_s}$, we know $W_{k_s} \cap E'_{s}=\varnothing$. Now note $\forall s \leq t$ if $s'> s$ then $E'_s \supseteq E_{s'}$ and so $W_{k_s} \cap E_{s'}=\varnothing$ as well, i.e. $k_{s'} \neq k_s$. Hence, every element of the sequence $k_0,...k_{t+1}$ is distinct. Contradiction. Thus $m_i$ has at most $2t+1$ switches after saying `Yes.'
\\
Now we show $\sigma_{m_i}^{-1}(\text{Yes})=\textbf{C}(P)$. Suppose $w \in \textbf{C}(P)$. Then $\exists k_w \leq t$ so that $w \in W_{k_w}=O_{k_w} \backslash O'_{k_w}$. Hence $\exists E_w \in \mathcal{E}_i(w)$ such that $E_w \subseteq O_{k_w}$. Furthermore $\forall E' \in \mathcal{E}_i(w)$, we have $W_{k_w} \cap E' \neq \varnothing$ as $w \in E'$ and $w \in W_{k_w}$. Thus, it follows $\forall E' \in \mathcal{E}_i(w)$ if $E' \subseteq E_w$ then $m_i(E')=\text{Yes}$, i.e. $w \in \sigma_{m_i}^{-1}(\text{Yes})$. On the other hand, suppose that $w \notin \textbf{C}(P)$. Take arbitrary $k \leq t$. We have that either $w \notin O_{k}$ or $w \in O'_{k}$. In case $w \notin O_{k}$ then $\forall E' \in \mathcal{E}_i(w)$ we know $E' \not \subseteq O_{k}$. Since $\mathcal{E}_i(w)$ is non-empty, we can pick any $E^*_k \in \mathcal{E}_i(w)$ so that $\forall E' \in \mathcal{E}_i(w)$ if $E' \subseteq E^*_{k}$ then $E' \not \subseteq O_k$. Alternatively, in case $w \in O'_{k}$, we can pick $E^*_k \in \mathcal{E}_i(w)$ so that $\forall E' \in \mathcal{E}_{i}(w)$ if $E' \subseteq E^*_k$ then $E' \subseteq O'_{k}$ and so $W_k \cap E'=(O_k \backslash O'_k) \cap E'=\varnothing$. Let $E^* \in \mathcal{E}_i(w)$ be an information state such that $E^* \subseteq \bigcap_{k \leq t} E^*_k$. By construction, $\forall E' \in \mathcal{E}_i(w)$ if $E' \subseteq E^*$  then $\forall k \leq t$ either $E' \not \subseteq O_k$ or $W_k \cap E'=\varnothing$ as $E' \subseteq E^*_k$, i.e. $m_i(E')=\text{No}$. Hence, $\sigma_{m_i}(w)=\text{No}$ and so $w \notin \sigma_{m_i}^{-1}(\text{Yes})$. 
\\
Thus, taking $n=2t+1$ and applying Theorem 3.0, we can conclude $\textbf{C}(P)$ is $(n+1)$-open in $\mathcal{T}_i$. $\square$
\newpage
\noindent \textbf{Theorem 7.2:} Suppose $\Omega$ is finite and each $n_i \geq n^*_{i,P}$. If $\textbf{C}(P)$ is non-empty then there is an attestation protocol $(s_i)_{i \in N}$  solving coordinated attack for $P$ whose success set is $\textbf{C}(P)$. Further, $(s_i)_{i \in N}$ is `welfare-maximizing' in the sense that $\textbf{L}(P)=\textbf{C}(P)$.\footnote[12]{This also means Lewis' definition of common inductive knowledge coincides with ours when each $n_i$ is sufficiently high.}
\\
\underline{Proof}
\\
By Lemma 7.1, $\textbf{C}(P)$ is $(n^*_{i,P}+1)$-open in each $\mathcal{T}_i$. Hence, $\textbf{C}(P)$ is $(n_i+1)$-open in each $\mathcal{T}_i$. Furthermore, since $\textbf{C}(P)$ is non-empty by assumption and is always a subset of $P$, it follows by Theorem 7.0 that there is an attestation protocol $(s_i)_{i \in N}$ solving coordinated attack for $P$ whose success set is $\textbf{C}(P)$. Additionally, since $\textbf{C}(P)$ is the success set of an attestation protocol which solves coordinated attack for $P$, we have that $\textbf{C}(P) \subseteq \textbf{L}(P)$. Thus, applying Lemma 7.0, we obtain $\textbf{L}(P)=\textbf{C}(P)$. $\square$

\section{Future Directions \& Conclusion}
In this paper, we gave a topological semantics for how common inductive knowledge can be generated by a shared witness. A natural extension would be to study how common inductive knowledge can be generated by a set of separate `local' witnesses for each agent. In the deductive case, Gonczarowski \& Moses 2024 have recently introduced a notion of asynchronous common knowledge which does essentially this; we believe our setup will be useful for generalizing their notion to an inductive context. Future work also ought to analyze how our logic changes when agents have unbounded switching tolerances. While we omitted answering this question in the present paper for sake of exposition, we expect our results will by and large carry over when dealing with arbitrary ordinals. Yet another extension of our logic worth pursuing is to model learners who receive noisy data. Genin \& Kelly 2017 examine how one might do topological learning theory in statistical settings and thus provide a potential path forward.
\\
\\
While we demonstrated that the proof system presented in this paper is sound with respect to the proposed semantics, a corresponding completeness result has not yet been obtained. We conjecture completeness holds, although it is possible that additional axioms or rules of inference will be necessary to fully capture our semantics. Regardless, any completeness result will have to confront a central technical difficulty stemming from the presence of a non-normal modality in our language-- namely, that of `having inductive reason to believe.' This operator fails to satisfy the distribution axiom and places standard completeness techniques for normal modal logics out of reach. A natural first step is therefore to restrict attention to a normal fragment of our language. One might then aim to establish completeness for that fragment by translating our semantics into an appropriate class of Kripke frames.
\\
\\
Lastly, we showed that our semantics lends itself to finding `attestation protocols' which solve an inductive variant of the well-known coordinated attack problem. The solutions to this problem ensure agents are coordinated in whether they converge on correctly attesting a proposition $P$ is true or whether they converge on deferring judgement. We hope future work finds such attestation protocols useful for designing novel and practical distributed consensus algorithms. For instance, imagine an `attestation aggregator' will be listening to each agent's output but is worried that some agents may be dishonest or faulty. If less than half of agents fall into this category then the aggregator can tell agents to follow an attestation protocol which solves coordinated attack for $P$ and, via majority vote, guarantee that it converges to attesting  $P$ is true if and only if all honest agents converge to correctly attesting $P$ is true. To conclude, the logic being developed here provides a fruitful foundation for reasoning about consensus in inductive settings and we are excited to see further progress along the directions outlined above.

\section{Bibliography}

\nocite{*}
\bibliographystyle{eptcs}
\bibliography{generic}

@BOOK{bookA,
    KEY       = "DL",
    AUTHOR    = "Lewis, D.",
    TITLE     = "Convention: A Philosophical Study",
    YEAR      = "1969",
    PUBLISHER = "Harvard University Press"
}

@BOOK{bookB,
    KEY       = "CS",
    AUTHOR    = "Cubitt, R. and Sugden, R.",
    TITLE     = "Common Knowledge, Salience, and Convention: A Reconstruction of David Lewis’ Game Theory",
    YEAR      = "2003",
    PUBLISHER = "Economics and Philosophy",
    VOLUME    = "19",
    DOI       = "https://doi.org/10.1017/S0266267103001123"
}

@BOOK{bookC,
    KEY       = "KK",
    AUTHOR    = "Kelly, K.",
    TITLE     = "The Logic of Reliable Inquiry",
    YEAR      = "1996",
    PUBLISHER = "Oxford University Press"
}

@BOOK{bookD,
    KEY       = "KK",
    AUTHOR    = "Kelly, K.",
    TITLE     = "Justification as Truth-Finding Efficiency: How Ockham’s Razor Works",
    YEAR      = "2004",
    PUBLISHER = "Minds and Machines",
    VOLUME    = "14",
    DOI       = "https://doi.org/10.1023/B:MIND.0000045993.31233.63"
}

@BOOK{bookE,
    KEY       = "HP",
    AUTHOR    = "Putnam, H.",
    TITLE     = "Trial and Error Predicates and the Solution to a Problem of Mostowski",
    YEAR      = "1965",
    PUBLISHER = "Journal of Symbolic Logic",
    VOLUME    = "30",
    DOI       = "https://doi.org/10.2307/2270581"
}

@BOOK{bookF,
    KEY       = "AT",
    AUTHOR    = "Tarski, A.",
    TITLE     = "A Lattice‑Theoretical Fixpoint Theorem and Its Applications",
    YEAR      = "1955",
    PUBLISHER = "Pacific Journal of Mathematics",
    VOLUME    = "5",
    DOI       = "https://doi.org/10.2140/pjm.1955.5.285"
}

@BOOK{bookG,
    KEY       = "CC",
    AUTHOR    = "Cousot, P. and Cousot, R.",
    TITLE     = "Constructive Versions of Tarski’s Fixed Point Theorems",
    YEAR      = "1979",
    PUBLISHER = "Pacific Journal of Mathematics",
    VOLUME    = "82",
    DOI       = "https://doi.org/10.2140/pjm.1979.82.43"
}

@BOOK{bookH,
    KEY       = "HM",
    AUTHOR    = "Halpern, J. and Moses, Y.",
    TITLE     = "Knowledge and Common Knowledge in a Distributed Environment",
    YEAR      = "1990",
    PUBLISHER = "Journal of the ACM",
    VOLUME    = "37",
    DOI       = "https://doi.org/10.1145/79147.79161"
}

@BOOK{bookI,
    KEY       = "GM",
    AUTHOR    = "Gonczarowski, Y. and Moses, Y.",
    TITLE     = "Common Knowledge, Regained",
    YEAR      = "2024",
    PUBLISHER = "Proceedings of the 25th ACM Conference on Economics and Computation (EC 2024)",
    DOI       = "https://doi.org/10.1145/3670865.3673442"
}

@BOOK{bookJ,
    KEY       = "KG",
    AUTHOR    = "Genin, K. and Kelly, K.",
    TITLE     = "The Topology of Statistical Inquiry",
    YEAR      = "2017",
    PUBLISHER = "Proceedings of the 16th Conference on Theoretical Aspects of Rationality and Knowledge (TARK 2017)",
    DOI       = "https://doi.org/10.4204/EPTCS.251.17"
}
\end{document}